\documentclass[a4paper,11pt]{article}
\pdfoutput=1 

\usepackage{jcappub} 

\usepackage[T1]{fontenc} 
\usepackage{xcolor}

\usepackage{graphicx}  
\usepackage{dcolumn}   
\usepackage{bm}        
\usepackage{amssymb}  
\usepackage[caption=false]{subfig}
\usepackage{esdiff}
\usepackage{amsmath}
\usepackage{commath}
\usepackage{hyperref}
\usepackage{natbib}
\hypersetup{colorlinks=true,allcolors=blue}
\newcommand*\difff{\mathop{}\!\mathrm{d}}
\newcommand{\lambdabar}{{\mkern0.75mu\mathchar '26\mkern -9.75mu\lambda}}

\title{\boldmath Cosmic ray ensembles as signatures of ultra-high energy photons interacting with the solar magnetic field}

\emailAdd{dalvarez@ifj.edu.pl}
\collaboration{The CREDO Collaboration}

\author[a,b]{N.~Dhital,} 
\author[b]{P.~Homola,}
\author[b,c]{D.~Alvarez-Castillo,}
\author[b]{D.~Gora,}
\author[b]{H.~Wilczy\'{n}ski,}
\author[b]{K.~Almeida Cheminant,}
\author[d]{B. Poncyljusz,}
\author[e]{J. M\k{e}drala,}
\author[e]{G. Opi{\l}a,}
\author[b]{G.~Bhatta,} 
\author[f]{T.~Bretz,} 
\author[g]{A. \'{C}wik{\l}a,} 
\author[h]{A.R.~Duffy,} 
\author[i]{A. C. Gupta,} 
\author[j]{B.~Hnatyk,} 
\author[e,b]{P.~Jagoda,}
\author[k]{M.~Kasztelan,}
\author[b]{K.~Kopa\'{n}ski,}
\author[l]{P.~Kovacs,} 
\author[b]{M.~Krupinski,}
\author[c]{V.~Nazari,}
\author[m]{M.~Nied\'{z}wiecki,}
\author[e]{D.~Ostrog\'{o}rski,}
\author[e]{K. Rzecki,}
\author[m]{K.Smelcerz,}
\author[n]{K.~Smolek,} 
\author[b]{J.~Stasielak,}
\author[b]{O.~Sushchov,}
\author[o,p]{T.~Wibig,} 
\author[b]{K.~Wozniak,}
\author[q,r]{J.~Zamora-Saa,} 
\author[l]{Z.~Zimbor\'{a}s,}
\author[s]{and Arman Tursunov}

\affiliation[a]{Central Department of Physics, Tribhuvan University, Kirtipur 44613, Nepal}
\affiliation[b]{Institute of Nuclear Physics PAN, Cracow 31-342, Poland}
\affiliation[c]{Joint Institute for Nuclear Research, Dubna, Russia}
\affiliation[d]{University of Warsaw, 02-093 Warsaw, Poland}
\affiliation[e]{AGH University of Science and Technology, 30-059 Cracow, Poland}
\affiliation[f]{RWTH Aachen University, III. Physikalisches Institut A, Aachen, Germany}
\affiliation[g]{Cracow University of Technology, 31-155 Cracow, Poland}
\affiliation[h]{Centre for Astrophysics and Supercomputing, Swinburne University of Technology, Hawthorn, VIC 3122, Australia}
\affiliation[i]{Aryabhatta Research Institue of Observational Sciences (ARIES), Manora Peak, Nainital 263001, India}
\affiliation[j]{Astronomical Observatory of Taras Shevchenko National University of Kyiv, Kyiv 04053, Ukraine}
\affiliation[k]{National Centre for Nuclear Research, 05-400 Otwock-Swierk, Poland}
\affiliation[l]{Institute for Particle and Nuclear Physics,Wigner Research Centre for Physics, Hungarian Academy of Sciences, H-1525 Budapest, Hungary}
\affiliation[m]{Institute of Telecomputing, Faculty of Physics, Mathematics and Computer Science, Cracow University of Technology, 31-155 Cracow, Poland}
\affiliation[n]{Institute of Experimental and Applied Physics, Czech Technical University in Prague, Prague, Czech Republic}
\affiliation[o]{University of \L \'{o}d\'{z}, Faculty of Physics and Applied Informatics, 90-236 \L \'{o}d\'{z}, Poland} 
\affiliation[p]{Cosmic Ray Laboratory, Astrophysics Division, National Centre for Nuclear Research, 90-558 \L \'{o}d\'{z}, Poland}
\affiliation[q]{Universidad Andres Bello, Departamento de Ciencias Fisicas, Facultad de Ciencias Exactas, Avenida Republica 498, Santiago, Chile}
\affiliation[r]{Millennium Institute for Subatomic physics at high
energy frontier - SAPHIR, Fernandez Concha 700, Santiago, Chile}
\affiliation[s]{Research Centre for Theoretical Physics and Astrophysics, Institute of Physics, Silesian University in Opava,
 Bezručovo nám. 13, CZ-74601 Opava, Czech Republic}

\abstract{Propagation of ultra-high energy photons in the solar magnetosphere gives rise to cascades comprising thousands of photons.
We study the cascade development using Monte Carlo simulations and find that the photons in the cascades are 
spatially extended over millions of kilometers on the plane distant from the Sun by 1 AU.
We compare results from simulations which use two models of the solar magnetic field, and show that although signatures of such cascades are different for 
the models used, for practical detection purpose in the ground-based detectors, they are similar.}

\begin{document}
\maketitle
\flushbottom

\section{Introduction}
Detection of ultra-high energy (UHE) photons, that bear energies of EeV and beyond, will have a significant impact on the understanding of fundamental science.
As an example, dark matter (DM) searches up to the electroweak scale ($\sim 100 \mathrm{\ GeV}$) so far have not been able to produce conclusive evidence of DM particles
\cite{PhysRevLett.107.131302, Ahmed:2009zw, PhysRevLett.112.091303}.
For this reason, it becomes even more important to explore the mass regimes corresponding to the other natural scales-- the GUT ($\sim 10^{16} \mathrm{\ GeV}$) and 
the Planck ($\sim 10^{19} \mathrm{\ GeV}$) scales for potential DM candidates \cite{Garny2016, 1475-7516-2015-08-024}.
A common method in the DM search has been the indirect search, which relies on the detection of products of DM particle decay and annihilation.
Various proposed models of particle interactions predict that products of such interactions consist of UHE photons and standard model (SM) particles with a possibility of
other elementary particles which do not fit into the SM \cite{BHATTACHARJEE2000109}.
Detection of UHE photons will also help substantiate the Greisen--Zatsepin--Kuzmin (GZK) effect, a steepening of cosmic ray energy spectrum around $ 4\times 10^{19}\mathrm{\ eV}$
as a consequence of interaction of UHE cosmic rays (UHECRs) with cosmic microwave background radiation \cite{PhysRevLett.16.748, Zatsepin:1966jv}.
Widely used techniques of UHE photon detection rely on two main approaches; first, analyses based on parameters (e.g., the depth of maximum development of an extensive air shower, $X_\mathrm{max}$) from 
the reconstructed longitudinal profiles of development of
extensive air showers (EASs) initiated by UHE photons \cite{ABRAHAM2009399}, and the other based on observables derived from signal recorded by ground-based detector arrays
from the secondary particles of EASs \cite{ABRAHAM2008243}.
In principle, both approaches should be able to distinguish between photon- and hadron-initiated showers. The photon-initiated showers are expected to have deeper $X_\mathrm{max}$
compared to the hadron-initiated ones, and the particle contents for the two types of showers are expected to be different-- hadronic showers being more
muon-rich than the other.
The most up-to-date results from searches implementing these techniques have reported not only the
non-observation  of (significant) photon candidates in UHECR data, thus enabling us to place stringent upper limits on UHE photon fraction (flux) \cite{Savina2021,Rautenberg:2021vvt,ABRAHAM2008243, ABRAHAM2009399},
but also the observation of an excess of muons in data compared to what one would expect from simulations of hadronic showers \cite{PhysRevD.98.022002, PhysRevD.91.032003}.
Given such a discrepancy between the measurements and the results from simulations of hadronic showers, which is possibly due to the lack of complete understanding
of the physics at the UHE regime, it is very appealing to revisit also the UHE photon scenario but with a different approach.

The alternative approach presented in this paper is based on the electromagnetic cascading of UHE photons traversing regions nearby the Sun.
Simulation results from a study of such a cascading process were presented in \cite{1999astro.ph.11266B}, which give an expected  size of a footprint 
of core part of the cascade at the top of the Earth's atmosphere. The footprint is expected to be a highly prolate ellipse with a size of the order of a few kilometers.
In our simulations, we take into account the more accurate physics of cascade development and tracking of the cascade particles so that we are able to 
characterize the particle distribution better. 
The cascading process starts when a UHE photon experiences
solar magnetic field component transverse to the direction of its trajectory sufficiently large for magnetic pair production. The electron-positron pairs thus produced
undergo a magnetic bremsstrahlung process and emit photons as they propagate in the magnetic field.
Also, among the emitted photons, those with sufficiently high energy will undergo magnetic pair production
and repeat the process. As a consequence, a cascade comprising several thousand photons and several $e^{+}e^{-}$ develops in the region nearby the Sun.
Although deflections suffered by the $e^{+}e^{-}$ during their propagation are very small
when considered only within these regions, they give rise to an extended spatial distribution of cascade particles after propagating through the Sun-Earth distance
($\sim 1.5\times10^{11} \ \mathrm{m}$). 
For UHE photons heading towards the Earth through the regions in the Sun's vicinity, a unique particle distribution is expected as the cascade reaches the Earth.
Such a cascading of UHE photons can occur even in the presence of the geomagnetic field \cite{2005CoPhC.173...71H}. However, cascades produced in the geomagnetic field, which are
  called {\it preshowers}, comprise only few hundred particles
and have very narrow spatial distribution ($< \mathrm{1\ m}$). Due to this fact, they are practically indistinguishable from the cascades without the preshower effect unless they originate at much higher altitude or arrive at the Earth's atmosphere at near horizontal direction. In the following part of this paper, we refer to the Sun-initiated cascades as {\it super-preshowers} (SPSs) in light of similar development
mechanism as that of preshowers but with much larger number of secondary particles. For a recent review on the different aspects of SPSs and state-of-the-art studies we refer the reader to this work~\cite{Homola:2020odt}.

\section{Simulation} 
The treatment of most of the physics processes involved in the simulation of SPS development has been adopted from the PRESHOWER program \cite{2005CoPhC.173...71H}.
We have used the formalism for magnetic pair production from reference \cite{RevModPhys.38.626}. For $n_\mathrm{photons}$ UHE photons propagating through a magnetic field ($H$), 
the actual number of $e^{+} e^{-}$
pairs produced ($n_\mathrm{pairs}$) is given by,
\begin{equation}
n_\mathrm{pairs} = n_\mathrm{photons} \{ 1 - \exp\left[-\alpha \left(\chi\right) \difff l \right] \} \mathrm{,} \label{eqn:n_pairs}
\end{equation}
where $\difff l$ is the photon path length and $\alpha\left(\chi\right)$ is the photon attenuation coefficient, 
a function of parameter $\chi \equiv \frac{1}{2} \frac{h \nu}{m_{\mathrm{e}} c^{2}} \frac{H}{H_\mathrm{cr}} $, where
$H_\mathrm{cr} \equiv \frac {m_{\mathrm{e}}^{2} c^{3}}{e \hbar} = 4.414 \times 10^{13} \mathrm{\ G}$ is the natural quantum-mechanical measure of magnetic field strength.
In an ultra-relativistic
limit, if $H \ll H_\mathrm{cr}$, $\alpha\left(\chi\right)$ can be expressed as
\begin{equation}
\alpha\left(\chi\right) = \frac{1}{2} \frac{\alpha_\mathrm{em}}{\lambdabar_\mathrm{c}} \frac{H}{H_\mathrm{cr}} T\left(\chi\right) \mathrm{,}
\end{equation}even millions of kilometers
where ${\lambdabar_\mathrm{c}}$ is the Compton wavelength of the electron and $T\left(\chi\right)$ is a dimensionless auxiliary function which can be approximated by
\begin{equation}
T\left(\chi\right) \simeq \frac{0.16} {\chi} K_{1/3}^{2} \left(\frac{2}{3\chi}\right) \mathrm{,} and 
\end{equation}
where $K_{1/3}$ is a modified Bessel function.
Provided the path length under consideration ($\difff l$) is fairly small, Eq. (\ref{eqn:n_pairs}) can be expressed as a probability of conversion of UHE photon into $e^{+} e ^{-}$ pair
 ($ p_\mathrm{conv} $) within the interval $\difff l$. Thus, we have
\begin{equation}
p_\mathrm{conv} = 1 - \exp\left(-\alpha \left(\chi\right) \difff l\right) \simeq \alpha \left(\chi\right)\difff l \mathrm{,} \label{eqn:p_conv}
\end{equation}
which for a much larger distance $L$ takes the form,
\begin{equation}
P_\mathrm{conv} = 1 - \exp[ - \int\limits_{0}^{L} \alpha \left(\chi\right)\difff l] \mathrm{.} \label{eqn:P_conv}
\end{equation}
The probability of conversion of a UHE photon into $e^{+} e ^{-}$ pair is evaluated using Eq. (\ref{eqn:p_conv}).
Also, a fraction of energy carried by a pair-member ($\varepsilon$) is chosen from the distribution
\begin{equation}
\od{n}{\varepsilon} \approx \frac{\alpha_\mathrm{em} H}{\lambdabar_\mathrm{c}} \frac{\sqrt{3}}{9\pi \chi} \frac{[2+\varepsilon(1-\varepsilon)]} {\varepsilon(1-\varepsilon)}
K_{\frac{2}{3}}\left[\frac{1}{3 \chi \varepsilon(1-\varepsilon)}\right]
\end{equation}
following \cite{1983ApJ...273..761D}.

As the conversion probability of UHE photons to $e^+ e^-$ pairs, their trajectories and characteristics of magnetic bremsstrahlung radiation emitted thereof depend on 
the magnetic field experienced by these particles along their trajectories, it is important that we incorporate a realistic solar magnetic field model in our simulation.
Unsurprisingly, owing to the dynamic nature and complexity of the Sun's magnetic field, a model that can characterize the magnetic field completely is far from being achievable.
Thus, we proceed first with a simple dipole model of solar magnetic field and later with another analytical model called the dipole-quadrupole-current-sheet (DQCS) model 
\cite{dqcs}.
\begin{figure}%
    \centering
    \subfloat[Dipole field]{{\includegraphics[width=0.35\textwidth]{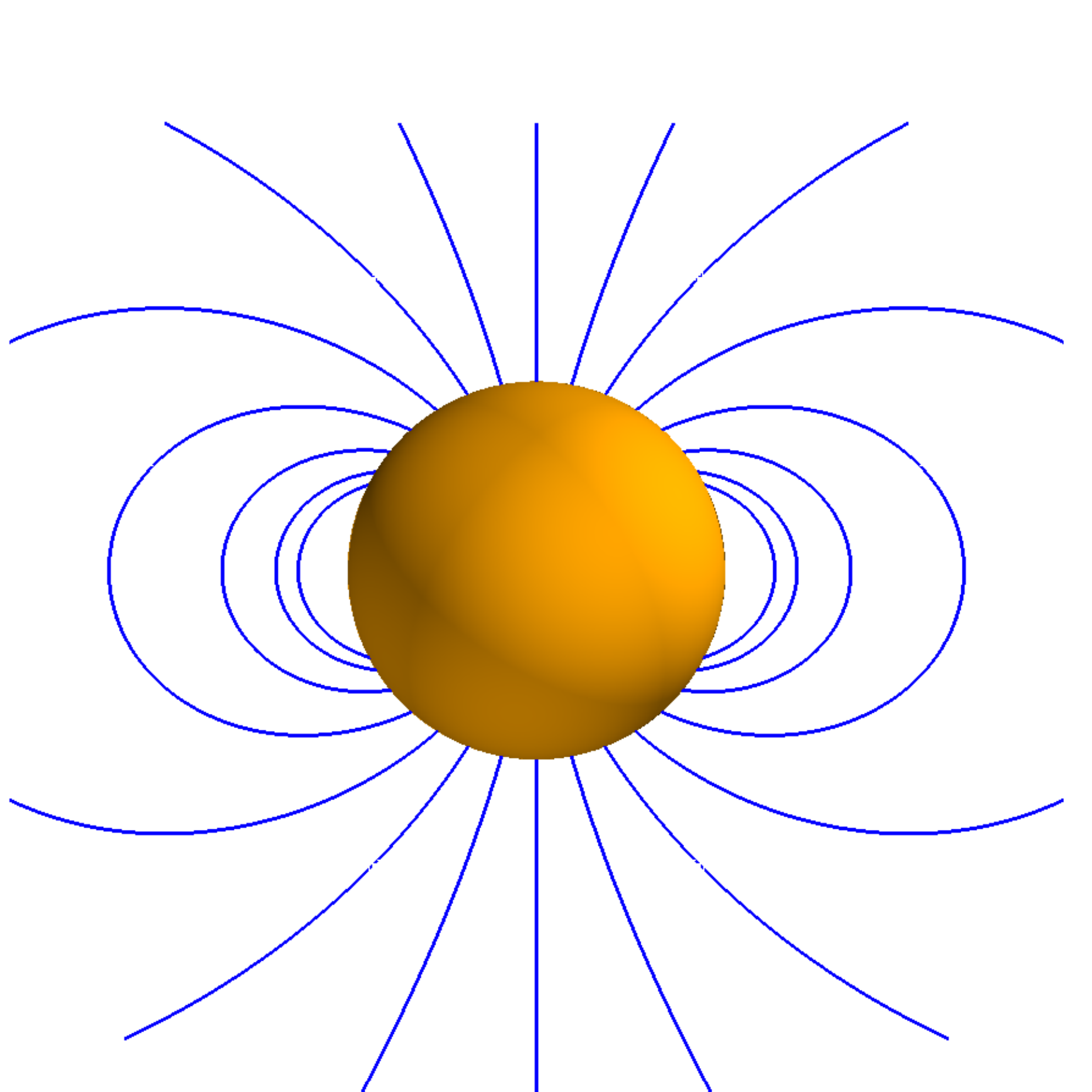}} }%
    \subfloat[DQCS field]{{\includegraphics[width=0.35\textwidth]{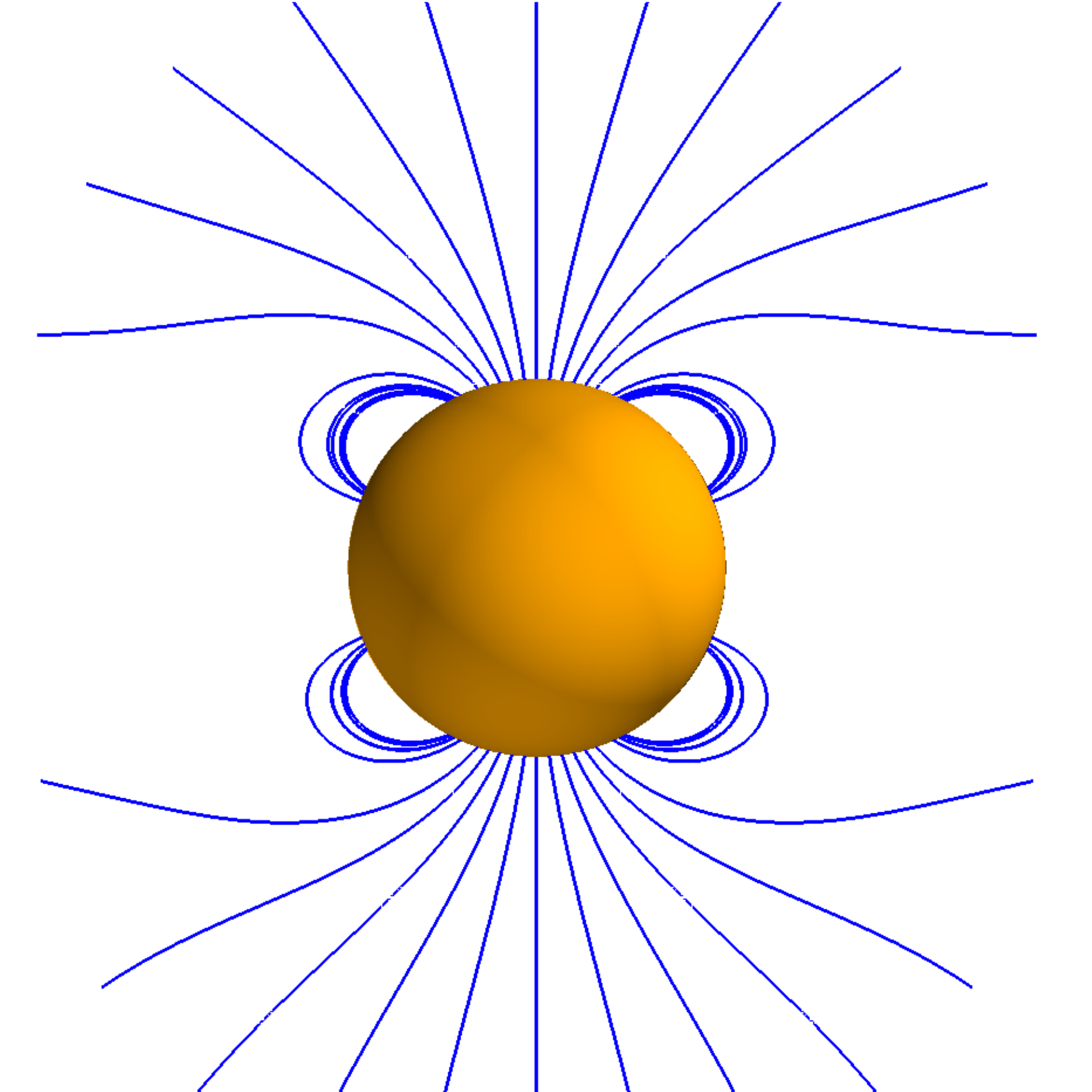} }}%
    \caption{Magnetic field configuration models used in the simulation.}%
    \label{fig:example}%
\end{figure}
For the dipole model, the magnetic moment of the dipole producing the field used in the simulation is $6.87 \times 10^{32} \mathrm{\ G \cdot cm^{3}}$.
Although this is not a very realistic model, using it for the solar magnetic field in our simulations allows us to study the effects of orientation of considered dipole
on the expected distribution of the particles in the SPS as they arrive at the top of the Earth's atmosphere. This will give us an idea of how the SPS development is affected by
an evolution of solar magnetic field for example over a solar cycle. 
In addition, it serves for a comparison to the results obtained from the other model. The DQCS model on the other hand, is more realistic and gives a more reasonable magnetic field
even in the interplanetary regions. Inclusion of this model in the simulation thus provides a more accurate tracking of $e^+ e^-$ on their way towards the Earth, and better treatment of  
magnetic bremsstrahlung processes.

We have introduced time and space tracking for particles in the cascade so that we obtain their arrival time distribution and lateral distribution
as they reach the top of the Earth's atmosphere. Given a particle with kinetic energy $E$ and charge $q$, propagating along the direction $\bm{\hat{v}}$ 
in a region defined by a magnetic field $\bm{B}$, the equation describing its motion as a function of time $t$ can be written as,
\begin{equation}
\diff{\bm{\hat{v}}\left(t\right)}{t} = \frac{q c^{2}}{E} \bm{\hat{v}} \times \bm{B} \mathrm{\quad .} \label{eqn1}
\end{equation}
The direction of propagation of a particle after it traverses a distance $\Delta s$ in time interval $\Delta t$ can be approximated by using a Taylor series expansion of 
$\bm{\hat{v}}\left(t + \Delta t\right)$ around $t$,
\begin{equation*}
\bm{\hat{v}}\left(t + \Delta t\right) \approx \bm{\hat{v}}\left(t\right) + \diff{\bm{\hat{v}}\left(t\right)}{t} \Delta t  \mathrm{,}
\end{equation*}
which takes the form
\begin{equation}
\bm{\hat{v}}\left(t + \Delta t\right)  \approx \bm{\hat{v}}\left(t\right) + \frac{q c^{2}}{E} \left(\bm{\hat{v}} \times \bm{B}\right) \Delta t \mathrm{,} \label{eqn2}
\end{equation}
after substituting $\diff{\bm{\hat{v}}\left(t\right)}{t}$ from Eq. (\ref{eqn1}).
We implement such a particle motion by choosing an appropriate $\Delta s$ which is split into two halves each equal to $\Delta s/2$. In the first half of the time interval
$\Delta t/2 = \Delta s / 2c$, the particle is propagated with the current direction vector which is then updated using Eq. (\ref{eqn2}) and is propagated with the new direction vector
for the latter half of the interval.

Using an expression for the spectral distribution of energy radiated by ultra-relativistic electron from \cite{Sokolov:1986nk}
\begin{eqnarray*}
f(y) = \frac{9\sqrt{3}}{8 \pi} \frac{y}{( 1 + \xi y )^3}\left\{\int\limits_{y}^{\infty} K_{\frac{5}{3}}(z)\difff z 
+\frac{(\xi y)^{2}}{1 + \xi y}  K_\frac{2}{3} (y)\right\}\mathrm{\quad}
\end{eqnarray*}
where parameter $\xi = \frac{3}{2} \frac{H_\perp}{H_\mathrm{cr}} \frac{E}{m_\mathrm{e} c^2}$, $E$ and $m_\mathrm{e}$ are energy and rest mass of electron respectively and $y$ is a 
function of emitted photon energy $h \nu$ defined by
\begin{equation*}
y\left(h \nu\right) = \frac{ h \nu}{\xi \left( E - h \nu \right)} \mathrm{,}
\end{equation*}
 {one can obtain the probability of emission of a bremsstrahlung photon from a sufficiently small path length $\difff l$. As has been derived in
\cite{2005CoPhC.173...71H}, the probability can be written as}
 {
\begin{equation}
P_\mathrm{brem}\left(B_\perp, E, h\nu, \difff l\right) = \difff l \int\limits_{0}^{E} I\left(B_\perp, E, h\nu\right) \frac{ \difff\left(h\nu\right)}{h\nu} \mathrm{,} 
\end{equation}
with
\begin{equation*}
I\left(B_\perp, E, h\nu\right) \equiv \frac{h\nu \difff N}{\difff \left(h \nu\right) \difff l} \mathrm{,}
\end{equation*}
where $\difff N$ is the number of photons with energy between $h\nu$ and $h \nu + \difff \left(h \nu \right)$ emitted over the path length $\difff l$. 
}

In addition, we have included the angular distribution of emitted synchrotron photons in our simulations. Since electrons are ultra-relativistic,
we take the half-opening angle of emitted synchrotron photons to be equal to $1/\gamma$, $\gamma$ being the Lorentz factor of the electron.
The azimuthal angle of emitted photon is randomly chosen from a uniform distribution {${\textit U}\left(0, 2\pi\right)$ \cite{SPS_Sim}. }

\section{Results}
We performed simulations for various representative cases of primary UHE photons traversing the Sun's vicinity on their way towards the Earth.
The solar magnetic field component transverse to the propagation direction of primary UHE photon has sufficiently large strength for pair production only in 
a small fraction of the path length close to the Sun. Emission of synchrotron photons from the $e^{+} e^{-}$ pair produced in this way also occurs mostly in the region near to the Sun.
Thus, almost the entire cascade development occurs in the close vicinity of the Sun.


\begin{figure*}[!bpht] 
\begin{center}$
\begin{array}{cc}
\includegraphics[width=0.5\textwidth]{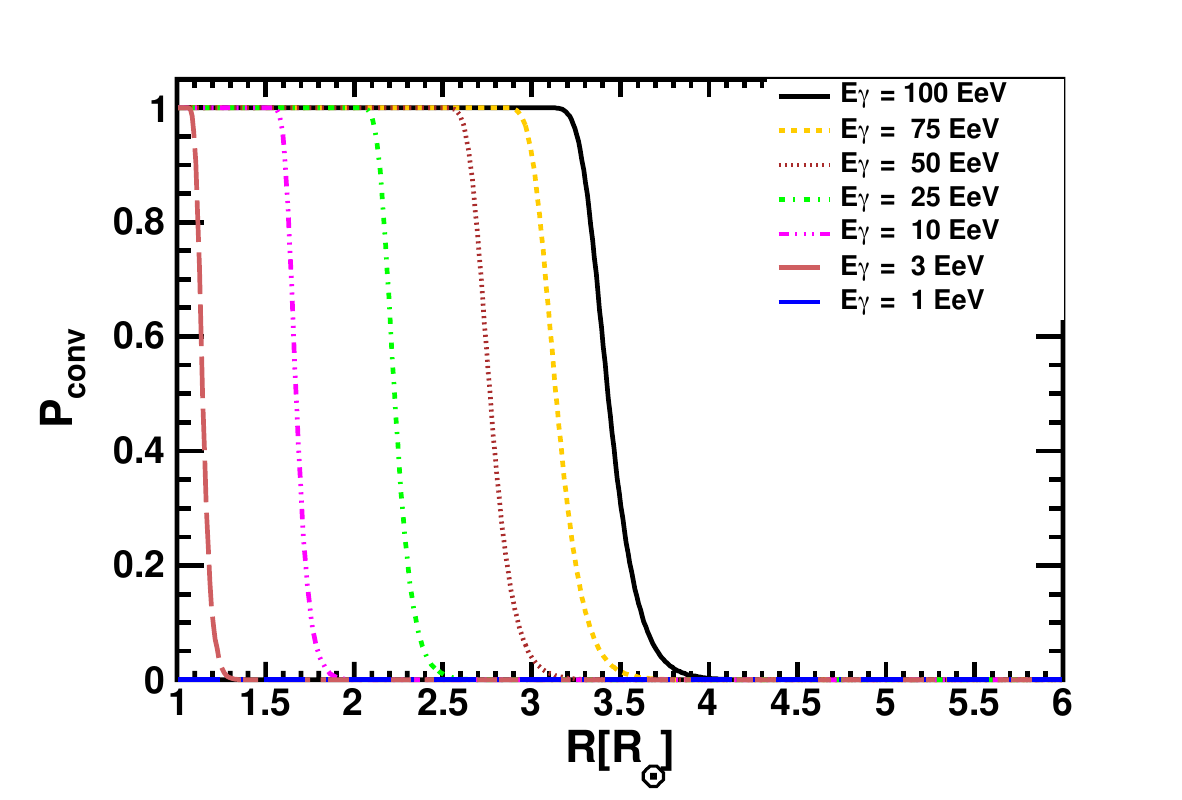} &\hspace{-0.5cm} \includegraphics[width=0.5\textwidth]{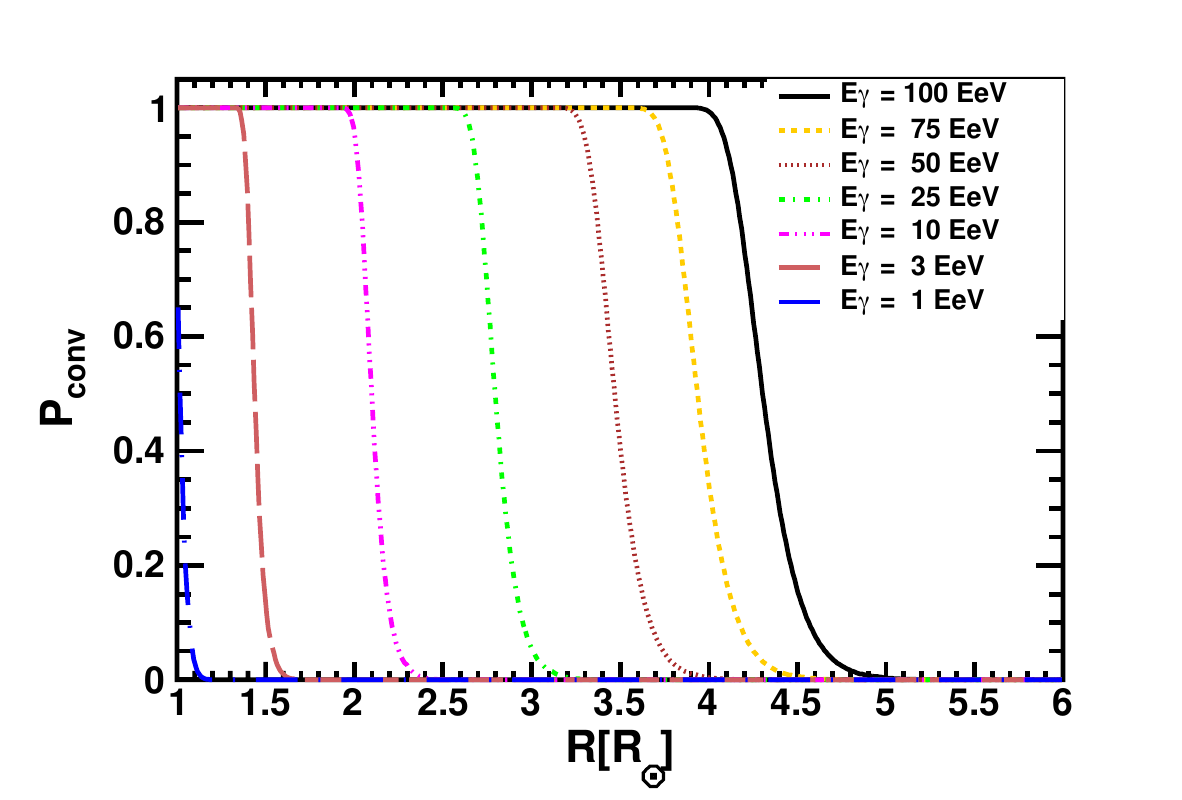}\\
\end{array}$
\end{center} 
\caption{Probability of magnetic pair production ($\gamma \rightarrow e^{+} e^{-}$) as a function of the impact parameter for UHE photons heading towards the Earth form the Sun's vicinity using the dipole magnetic field model.
\textit{Left panel:} Equatorial photon incidence.
\textit{Right Panel:} Polar photon incidence.}
\label{fig:conversion_p}
\end{figure*}

\begin{figure}
\centering
    \includegraphics[width = 0.7\textwidth]{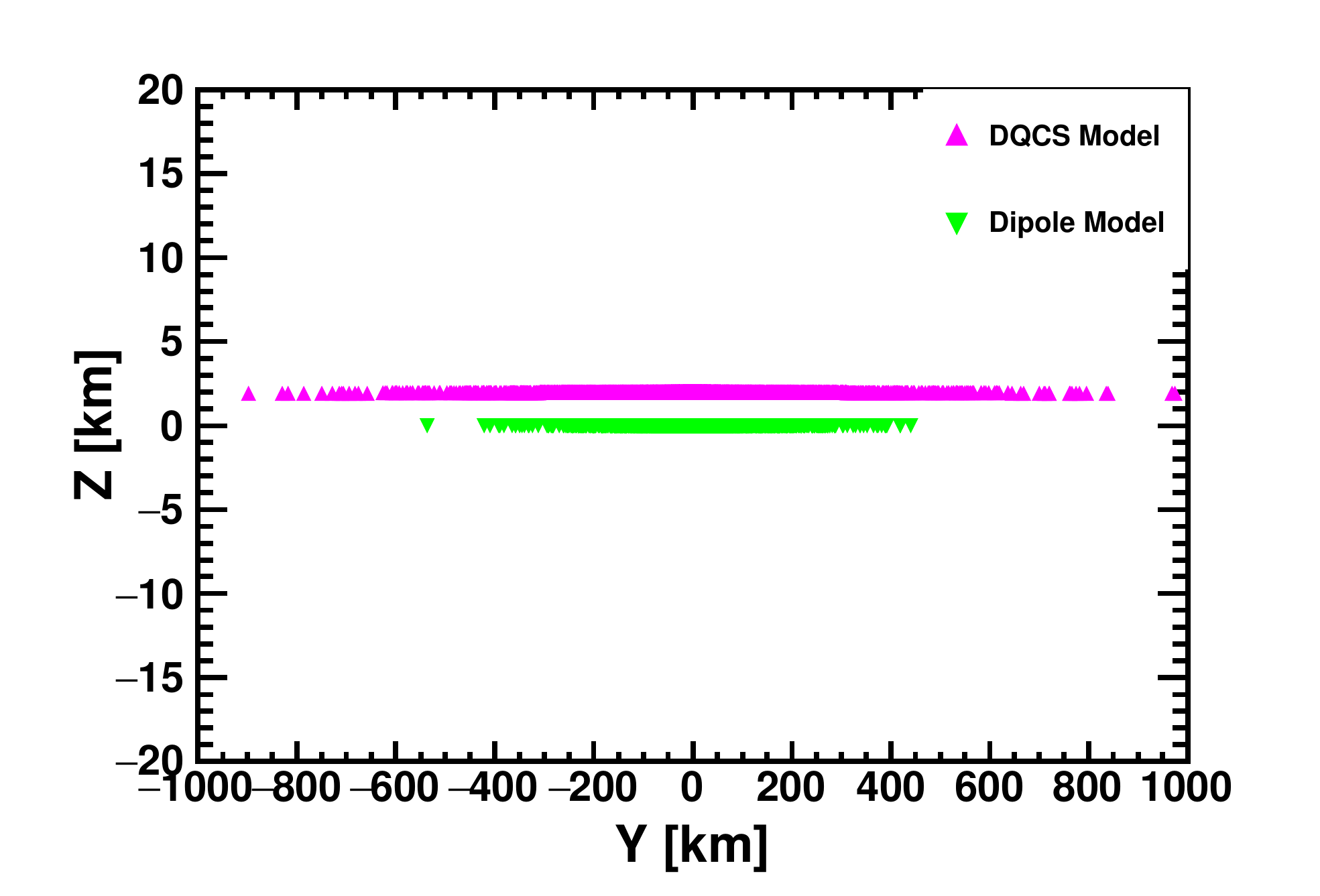}
    \includegraphics[width = 0.7\textwidth]{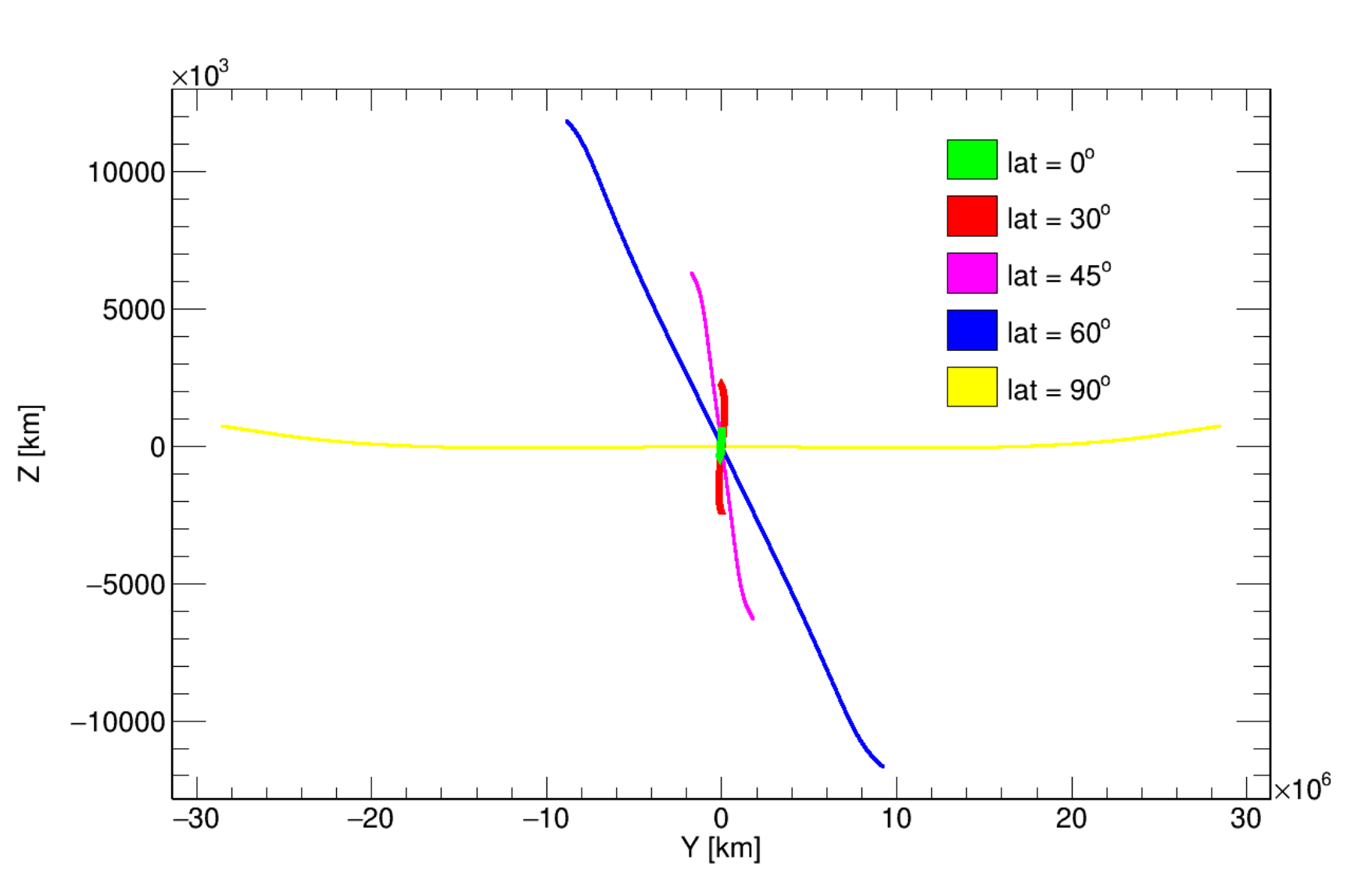}
\caption{Spatial distribution of photons with energies $> 10^{6} \mathrm{\ eV}$ arriving at the top of the atmosphere for an SPS produced by a 100 EeV photon. 
The primary photon is directed towards the Earth such that the position of the closest approach has several heliocentric latitudes: $0^\circ$, $30^\circ$, $45^\circ$, $60^\circ$, and $90^\circ$. The impact parameter is fixed with a value of 2$R_{\odot}$.
In the top panel, the presented distribution corresponds to the dipole model of the magnetic field of the Sun whereas the bottom panel displays the results for the DQCS model.
} \label{fig:signature}
\end{figure}
The electron and the positron, although travelling along slightly
different tracks, experience practically the same transverse magnetic field strength.
The electron and the positron are deviated
in opposite directions approximately in the same plane, when considered only in the small region where most of the cascade develops. 
The argument that their motion is approximately in the same plane comes from the fact that 
for the highly energetic $e^{+} e^{-}$ travelling in a magnetic field of which the strength typically is much less than a Gauss, 
the gyroradius of the motion is much larger than the 
length of the track where they experience this field. Synchrotron photons emitted from these ultra-relativistic electrons are highly beamed in the forward direction of the latter,
which gives rise to spatial distribution that has a very elongated footprint, when the cascade arrives at the top of the Earth's atmosphere.
The probability of conversion of a 100 EeV UHE photon propagating towards the Earth from the Sun's vicinity as a function of its impact parameter is 
shown on Fig.~\ref{fig:conversion_p} for equatorial and polar incidence. 
The conversion probability is close to unity for impact distance as far as {$4R_\odot$} for equatorial incidence from the Sun's center for a 100 EeV photon, 
which translates to the fact that despite a small solid angle
subtended by the Sun while viewed from the Earth, the effective solid angle relevant for SPS search is about 15 times larger at this energy. However, for lower energies around 10 EeV,
the conversion probability is close to unity as far as  $2R_\odot$, thus giving a region 3 times larger than the apparent size of the Sun viewed from the Earth.
For the case when a primary photon traverses a region very close to the Sun ($\sim 1 R_\odot$), conversion probability is close to 1 even for a 1 EeV photon. Similar conclusions apply to the photon polar incidence case, exhibiting slightly larger values for the corresponding impact parameter. 
Also, in Fig. \ref{fig:signature}, spatial distribution of photons for an example case is shown.
In the plot, $y = 0, z = 0$ corresponds to the
point at the top of the Earth's atmosphere where the UHE photon would have landed, had there been no interaction on its way. Positive $y$ and $z$ axes point towards
the East and the North directions respectively.
Although the particle distribution is dependent on the solar magnetic field model
used in the simulation as is evident in the figure, the nature of the particle distribution
(i.e., a very extended spatial distribution) holds for both models.
\begin{figure}[htpb!]%
    \centering
    \includegraphics[width = 0.8\textwidth]{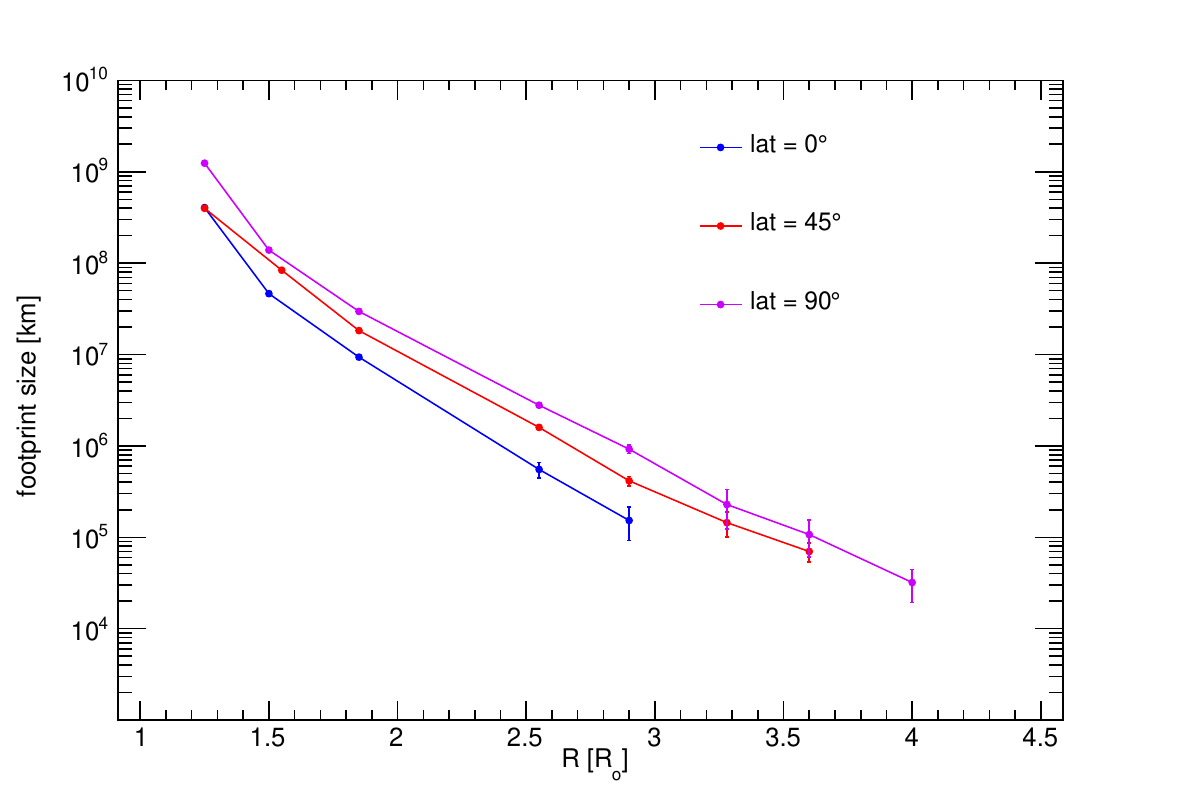}
    \includegraphics[width = 0.8\textwidth]{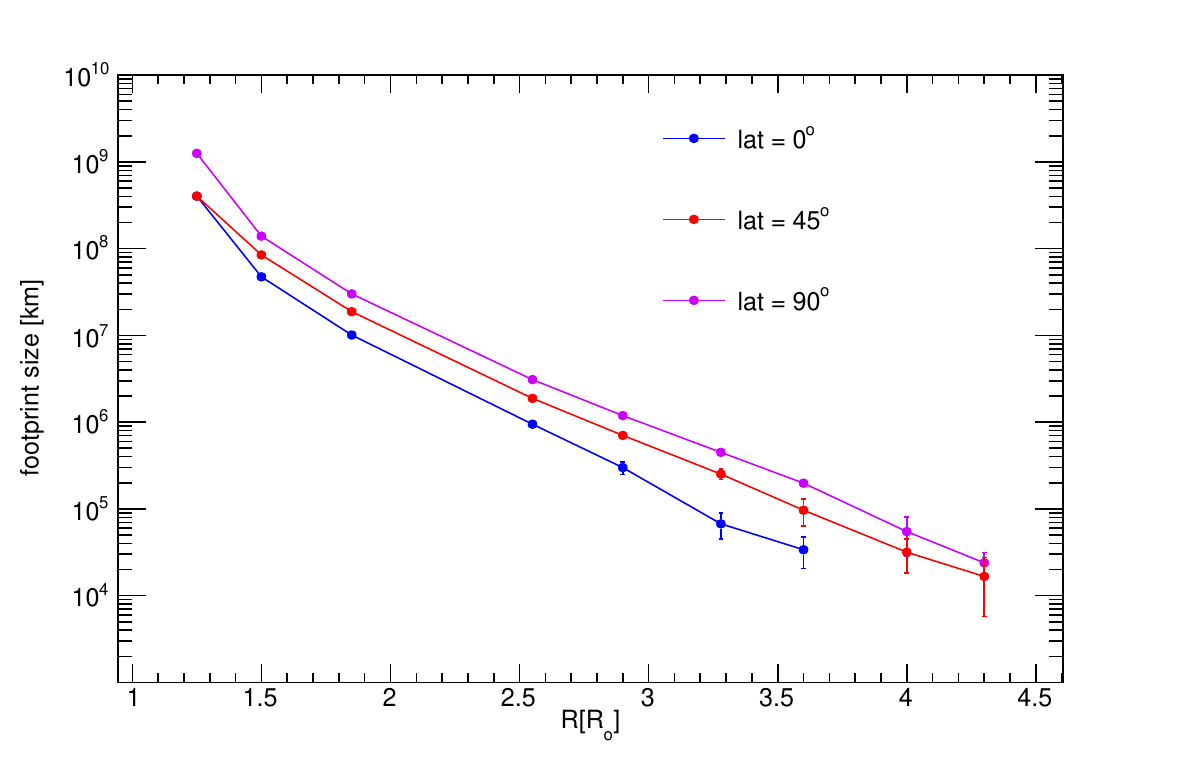}
    \caption{    
  Size of SPS footprint at a distance of 1 AU from the Sun as a function of impact parameter $R$ for a primary photon with an energy of 50 EeV (top panel) and 100 EeV (bottom panel). The SPS footprint size is defined as the spatial extent of photons with energies of 1 MeV and higher on the plane distant from the Sun by 1 AU.The values obtained in these plots correspond to the dipole magnetic field model of the Sun.}
\label{fig:foot_print_size}
\end{figure}
\begin{figure}[htpb!]%
\centering
   \includegraphics[width = 0.7\textwidth]{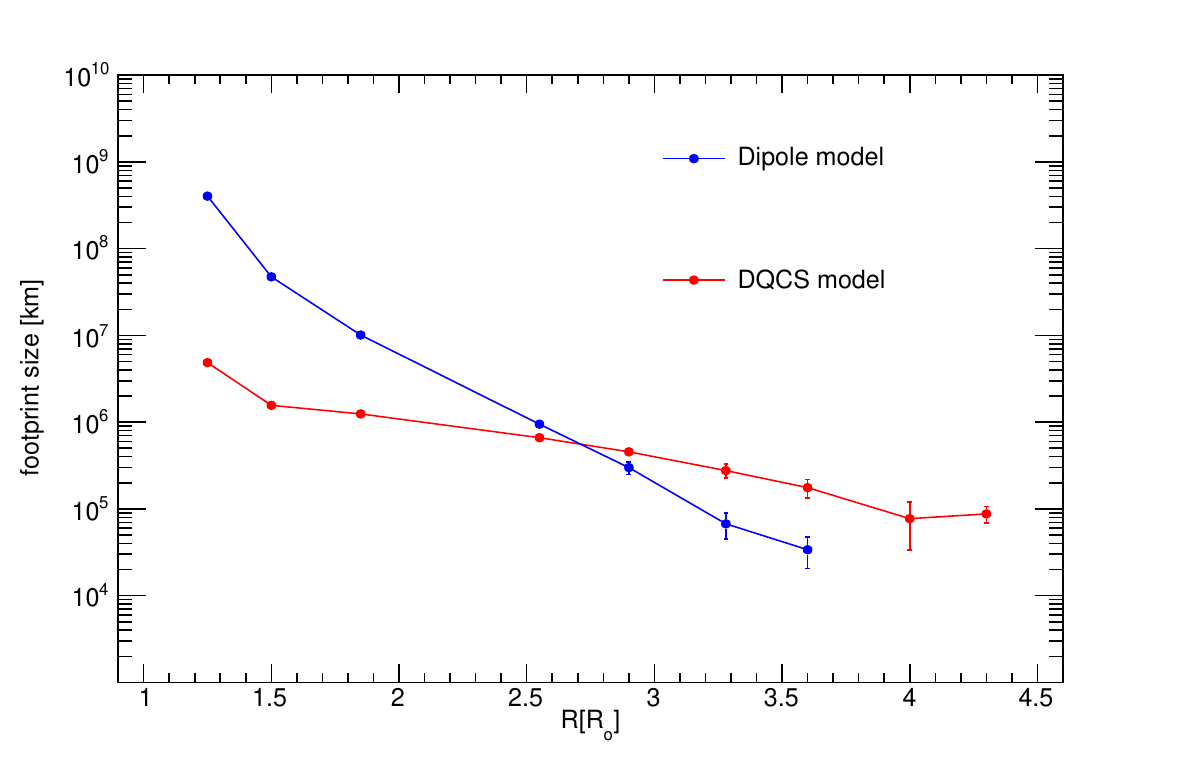}
   \includegraphics[width = 0.7\textwidth]{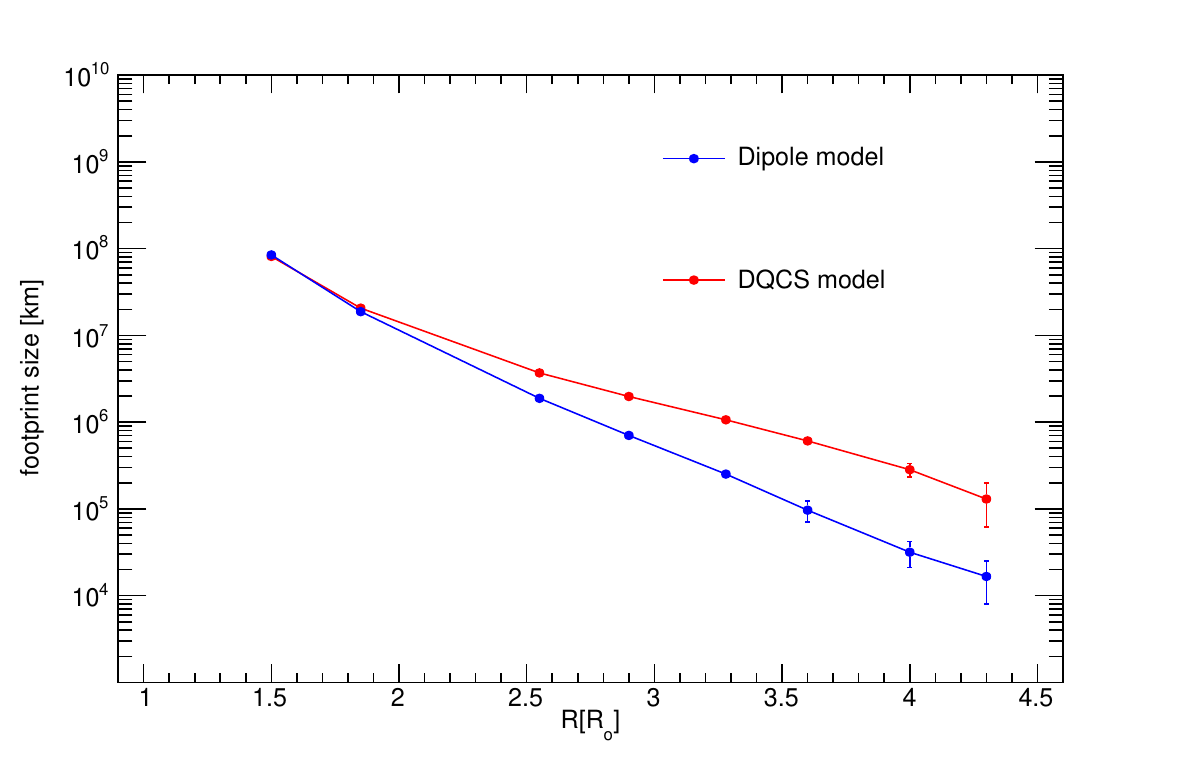}
\caption{Size of footprint for the DQCS and Dipole Model as a function of impact parameter $R$ for a primary photon with an energy  of 100 EeV with a) latitude $0^\circ$ (top panel) and b) latitude $45^\circ$ (bottom panel). The SPS footprint size is defined as the spatial extent of photons with energies of 1 MeV and higher on the plane distant from the Sun by 1 AU.
\label{fig:foot_print}
}
\end{figure}
\begin{figure}%
    \centering
    \includegraphics[width = 0.8\textwidth]{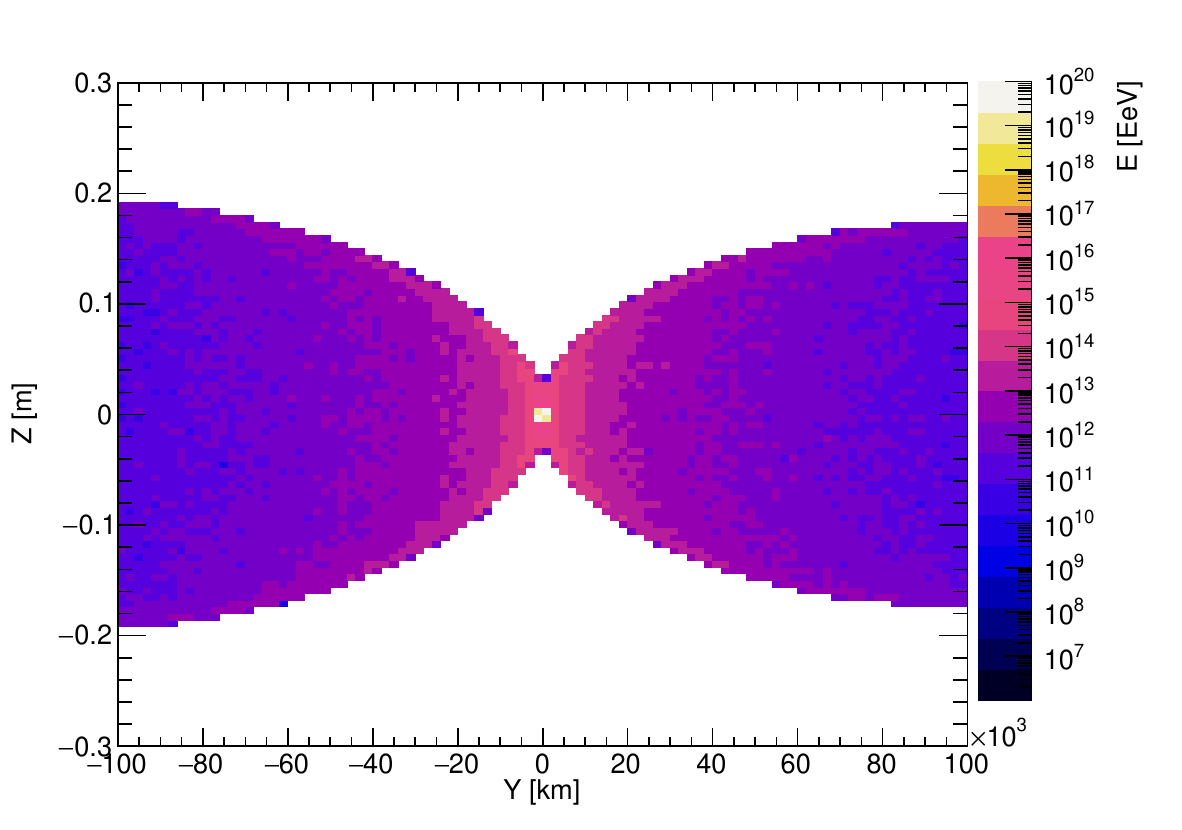}
    \caption{Distribution of energy of SPS photons arriving at the top of the atmosphere in a central region of an SPS produced by a 100 EeV photon. The primary photon is directed towards the Earth such that the position of the closest approach has heliocentric latitude $0^\circ$, and its impact parameter is $3 R_\mathrm{\odot}$. Note the difference in the scales along $y$ and $z$ axes.
    }\label{fig:distr}
\end{figure}
\begin{figure}%
    \centering
    \includegraphics[width = 0.7\textwidth]{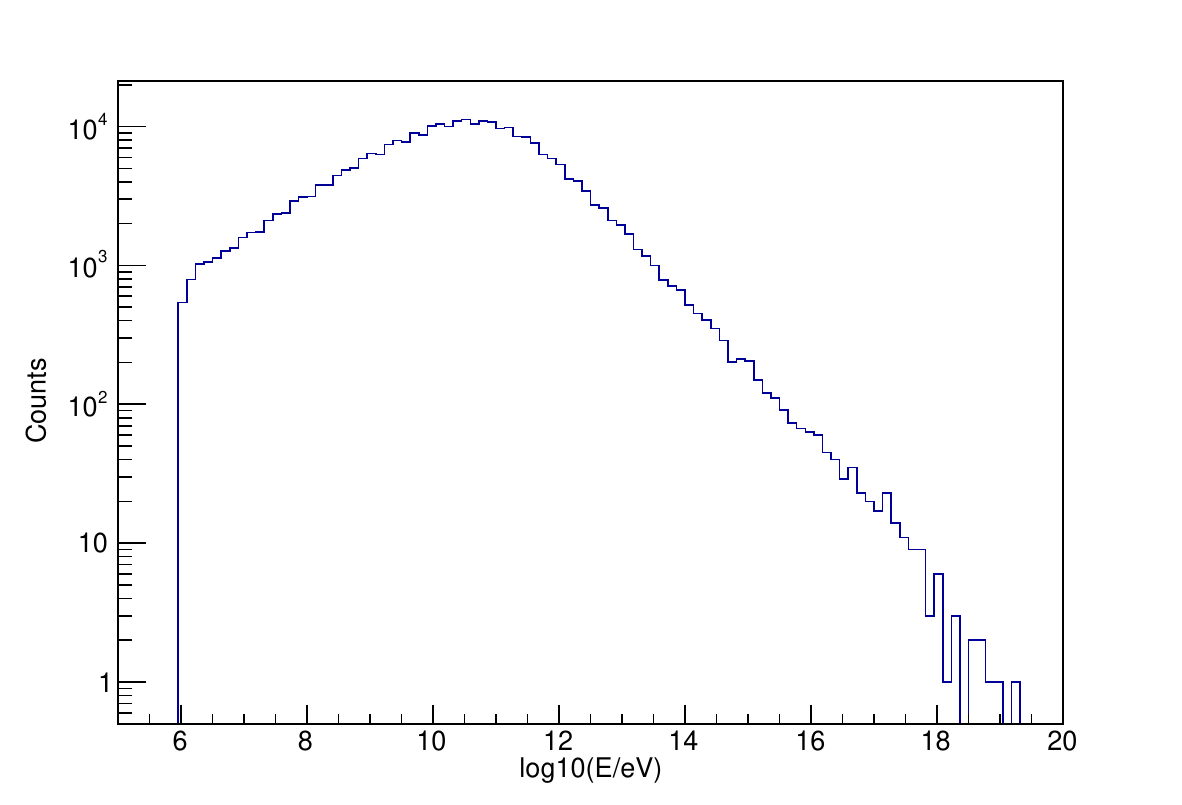}
    \caption{Energy distribution of SPS photons with energies larger than  $10^{6}~\mathrm{eV}$ for a SPS produced by a 100 EeV photon. Such a primary photon is directed towards the Earth and its impact parameter is 3$R_{\odot}$. 
}\label{fig:synch} 
\end{figure}
A salient feature of SPSs we observe in our simulation results is a very extended spatial distribution of cascade particles, apparently along a straight line, 
as the cascade reaches the top of the Earth's atmosphere. This extended footprint is a straightforward consequence of the deviation of electrons and positrons along their
tracks under the influence of (practically the same) solar magnetic field, and emission of highly forward-beamed synchrotron photons from them as they propagate towards
the Earth. In Fig. \ref{fig:foot_print_size}, dipole model SPS footprint sizes for 50 and 100 EeV photons heading towards the Earth from different directions are shown. SPS footprint size in plots \ref{fig:signature}, \ref{fig:foot_print_size}, \ref{fig:foot_print} is defined as the spatial extent of photons with energies of 1 MeV and higher on the plane distant from the Sun by 1 AU.
The plots are obtained from 1000 simulations each with the impact position of photons heading towards the Earth randomly chosen from a uniform two dimensional distribution around the Sun
such that the range of impact parameter is between $1 R_\odot$ and $5R_\odot$. Figure \ref{fig:foot_print} presents the results of both models for $0^\circ$ and $45^\circ$ latitudes.
In Fig. \ref{fig:distr}, the particle distribution at the top of the Earth's atmosphere weighted by particle energy for an example simulation is shown. The central region of the cascade comprises the most energetic photons. Figure \ref{fig:synch} shows the corresponding energy distribution of photons in the SPS cascade displayed in the previous figure.  An important implication of remarkably large sizes of SPSs demonstrated in Figs. \ref{fig:signature}, \ref{fig:foot_print_size}, and \ref{fig:foot_print} is that SPS tails might reach Earth even if the primary UHE photons which initiate these SPSs arrive from the directions much different from the direction of the Sun, i.e., practically from the sky hemisphere with the Sun in its center.
\subsection{Multi air shower footprints  at the ground level}

\begin{figure}
\includegraphics[width=1.0\textwidth]{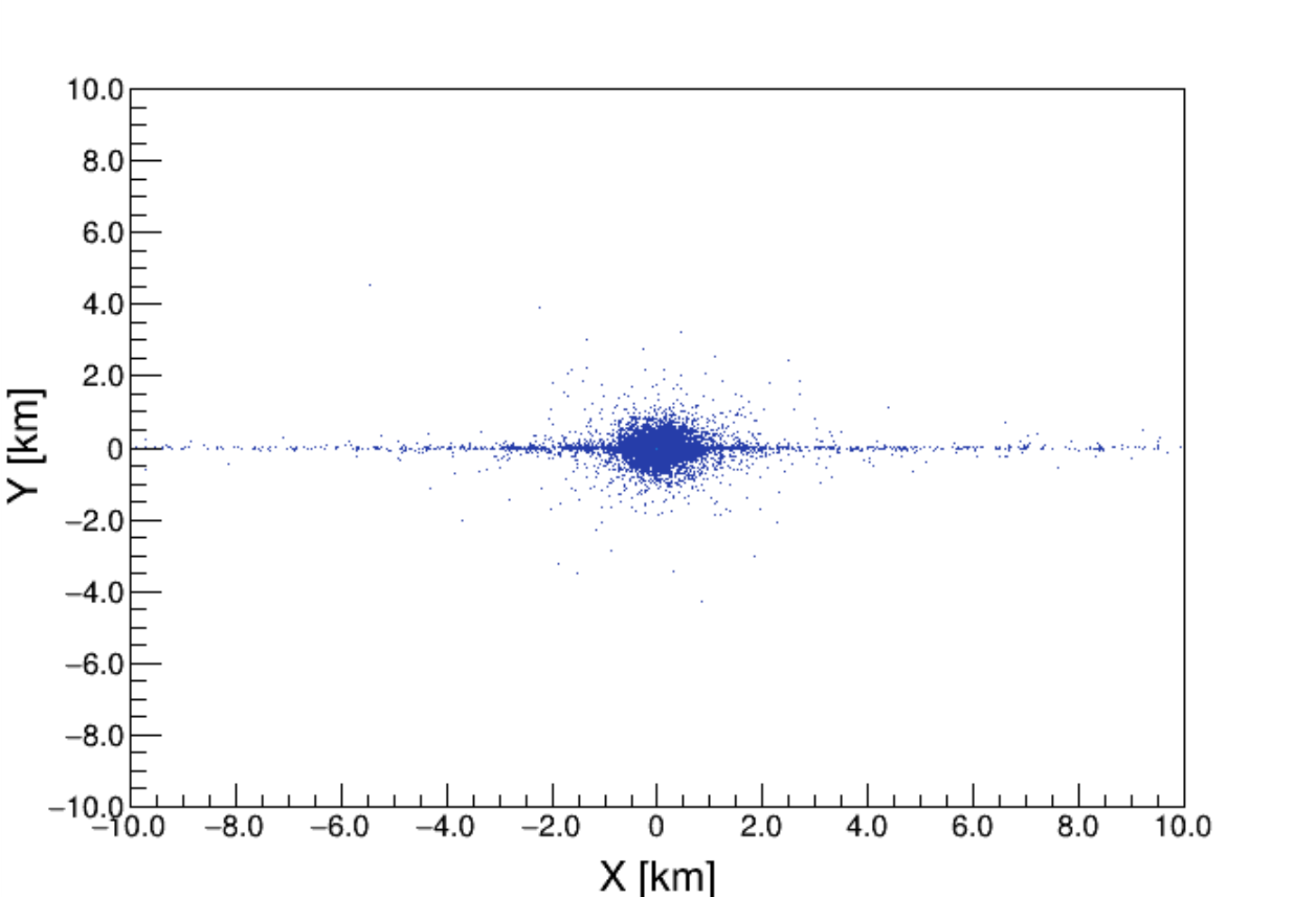}%
\begin{picture}(0,0)
\put(-180,40){\includegraphics[height=3.25cm]{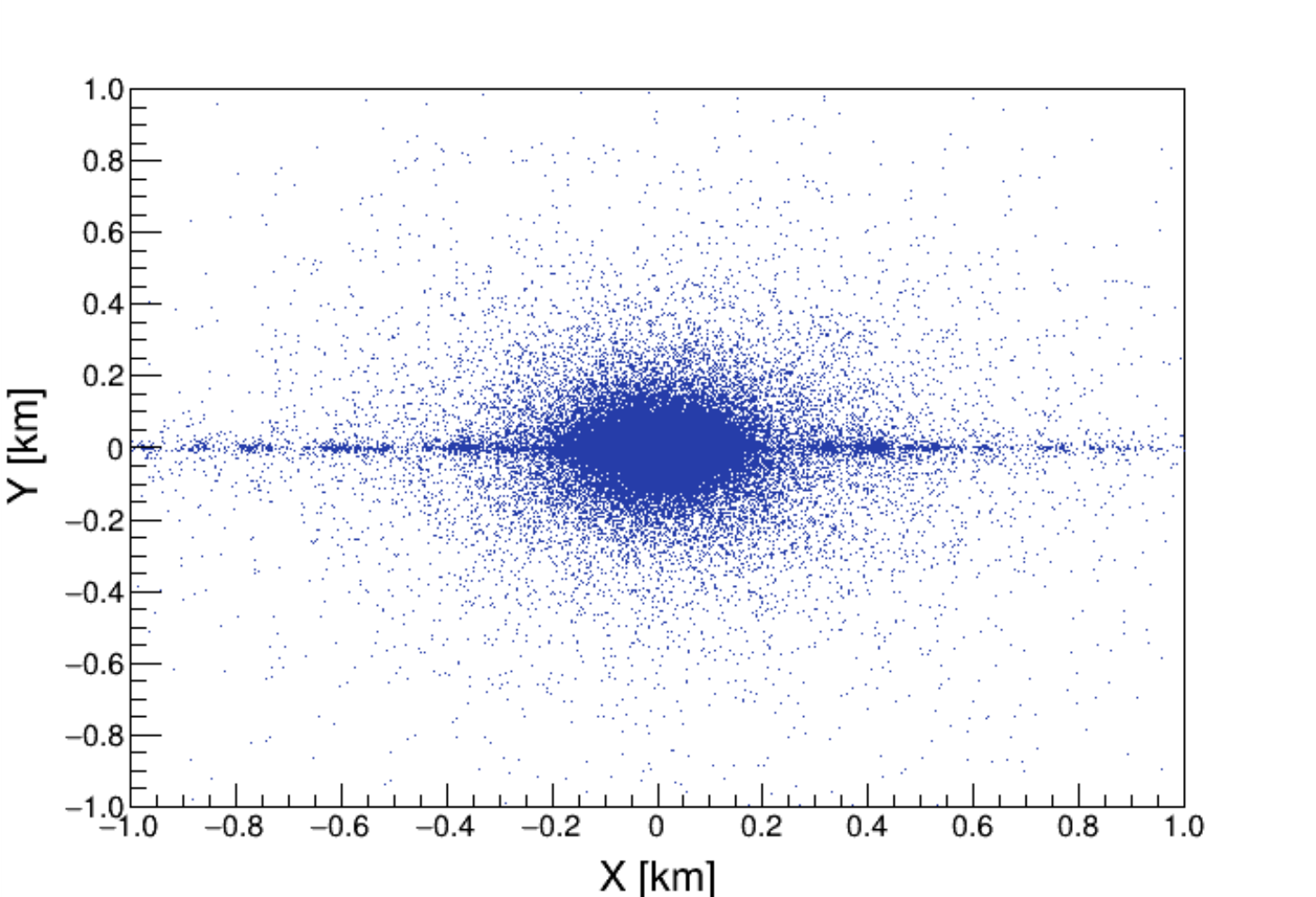}}
\end{picture}
\caption{The central regions of an example multi air shower particle distribution on ground generated by an SPS originated from a primary UHE photon of energy $10^{19}$ eV. The inset displays the core of the footprint in a smaller area.}
\label{fig:footprint}
\end{figure}

\begin{figure*}[!bpht]
\begin{center}$
\begin{array}{cc}
\includegraphics[width=0.45\textwidth]{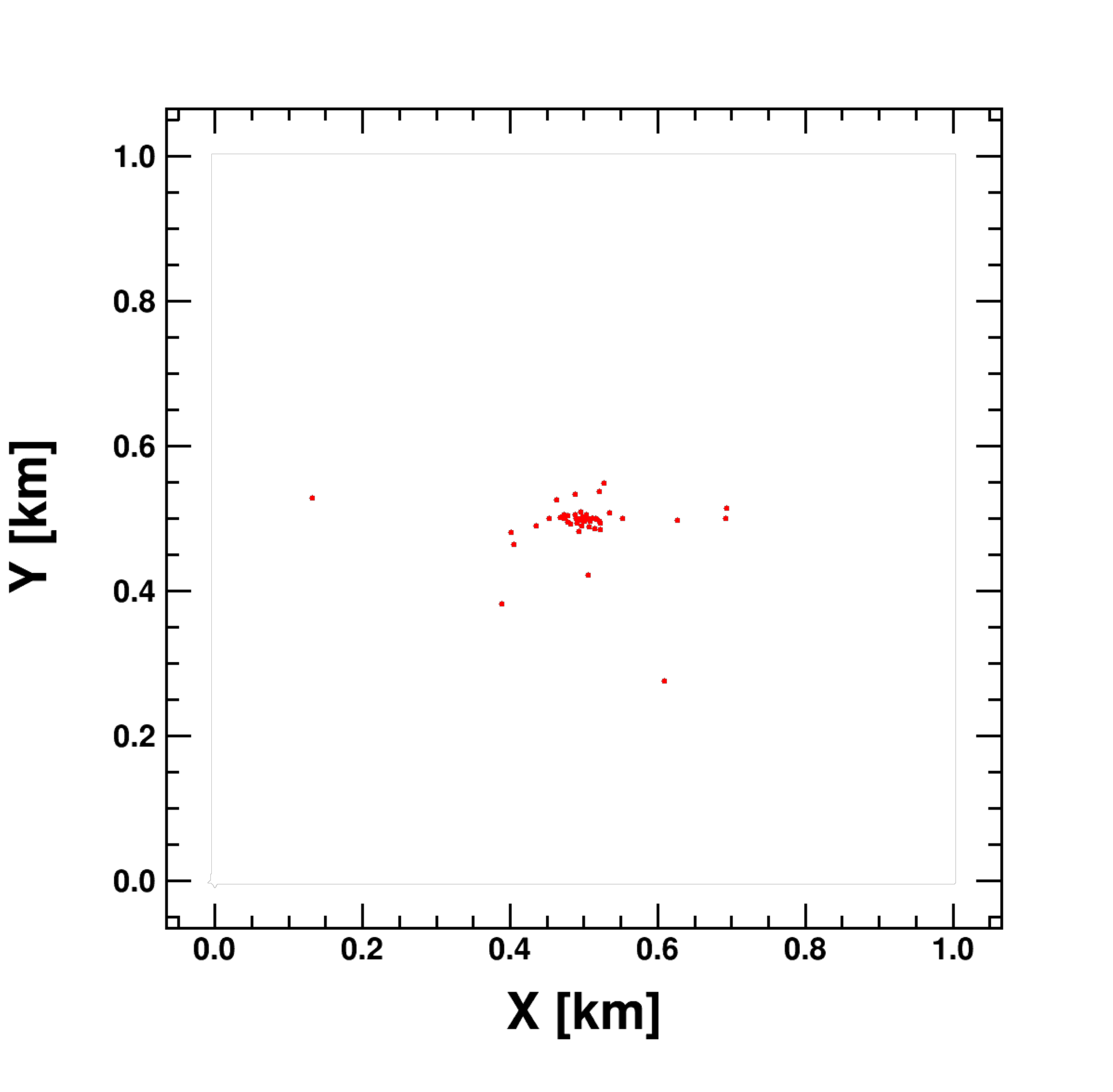} & \includegraphics[width=0.45\textwidth]{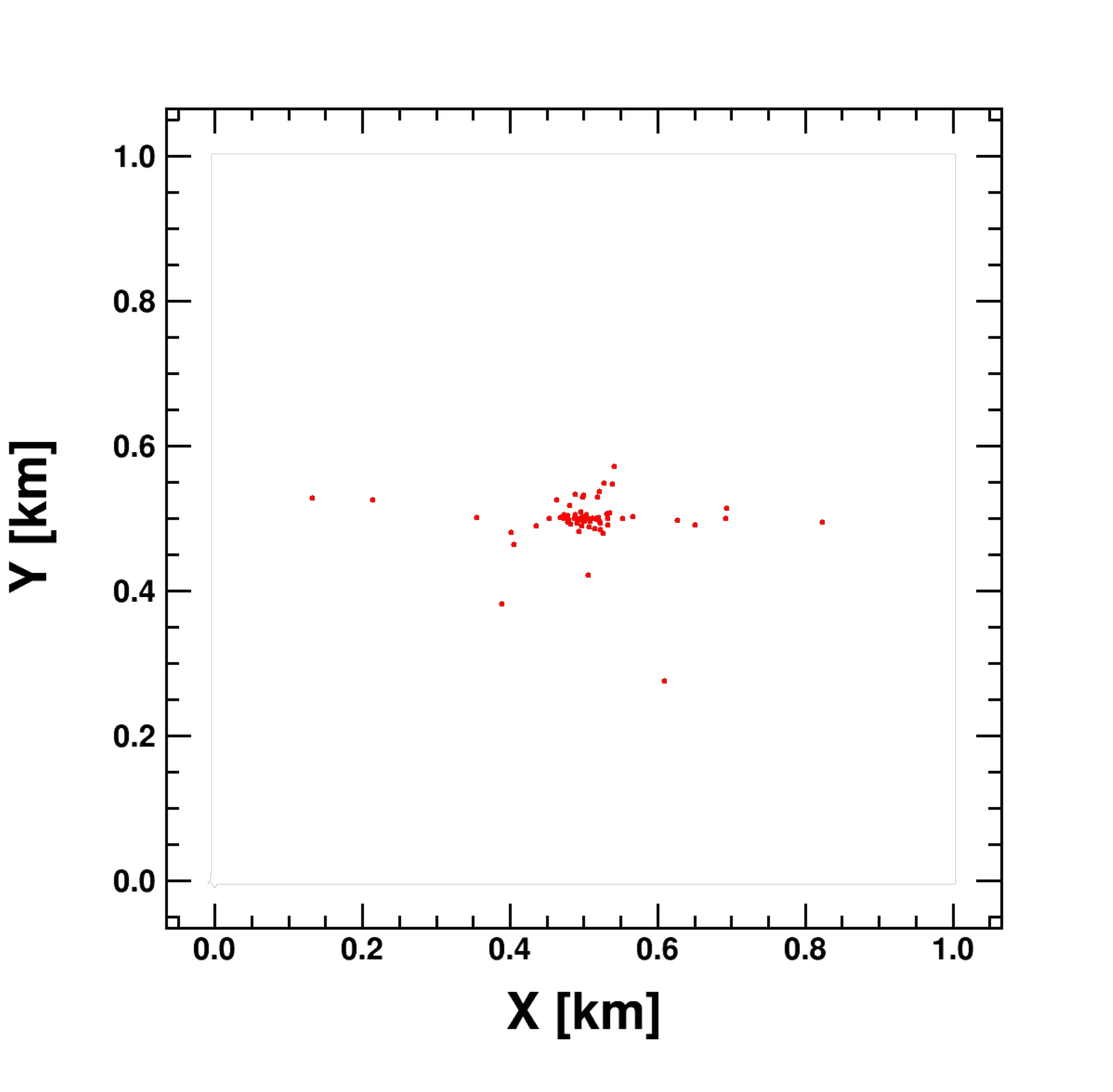}\\
\includegraphics[width=0.45\textwidth]{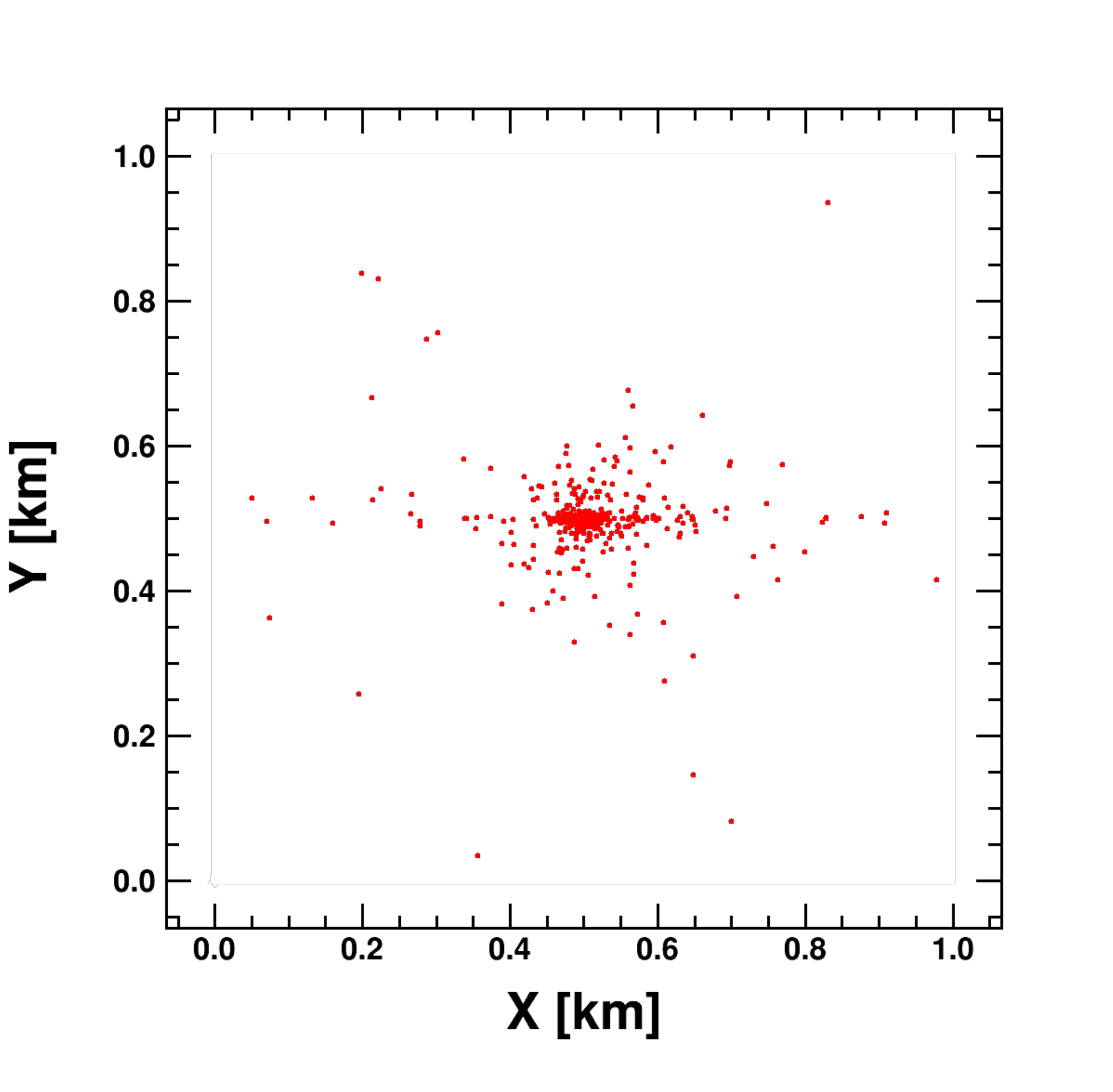} & \includegraphics[width=0.45\textwidth]{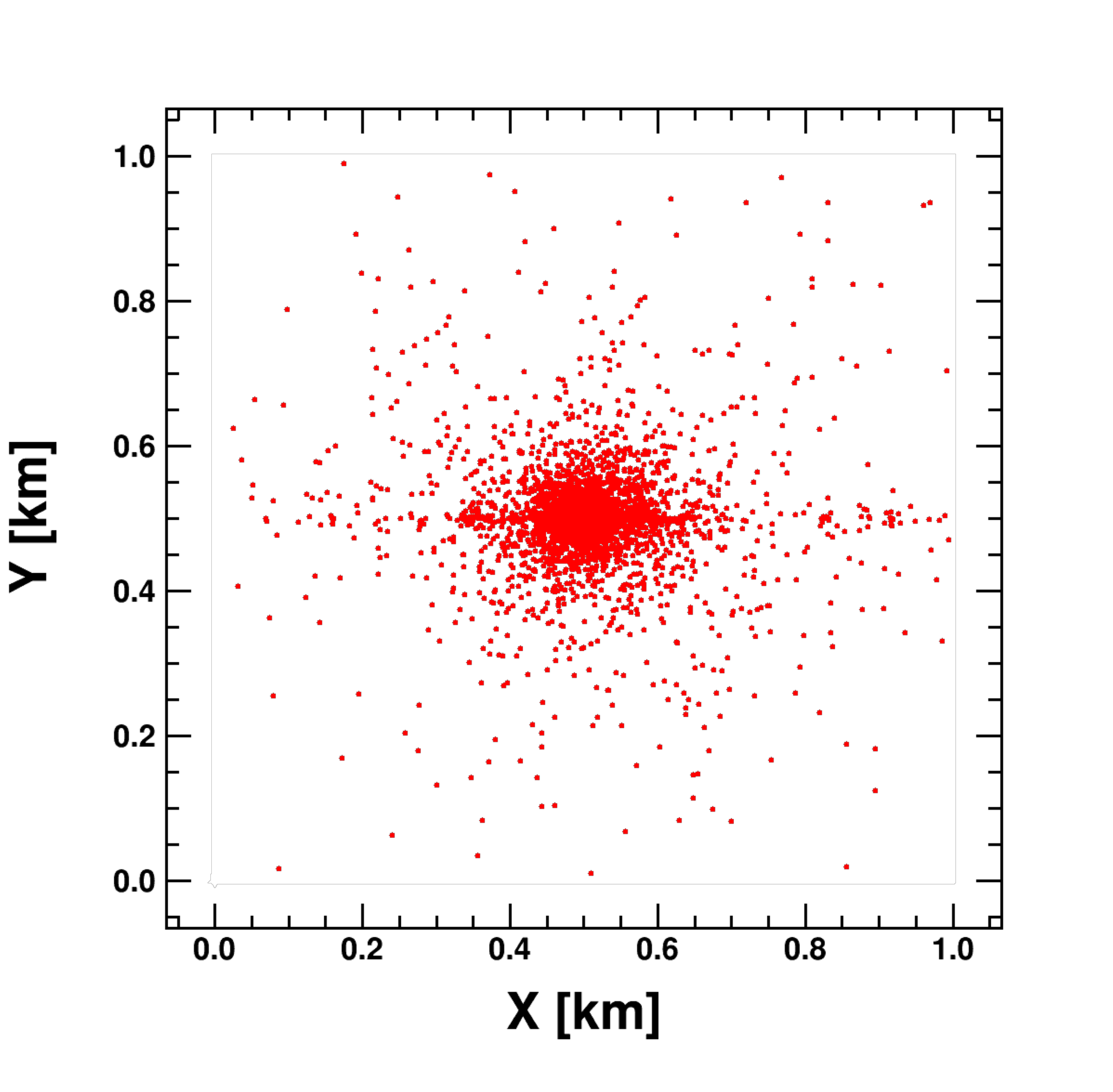}\\
\includegraphics[width=0.45\textwidth]{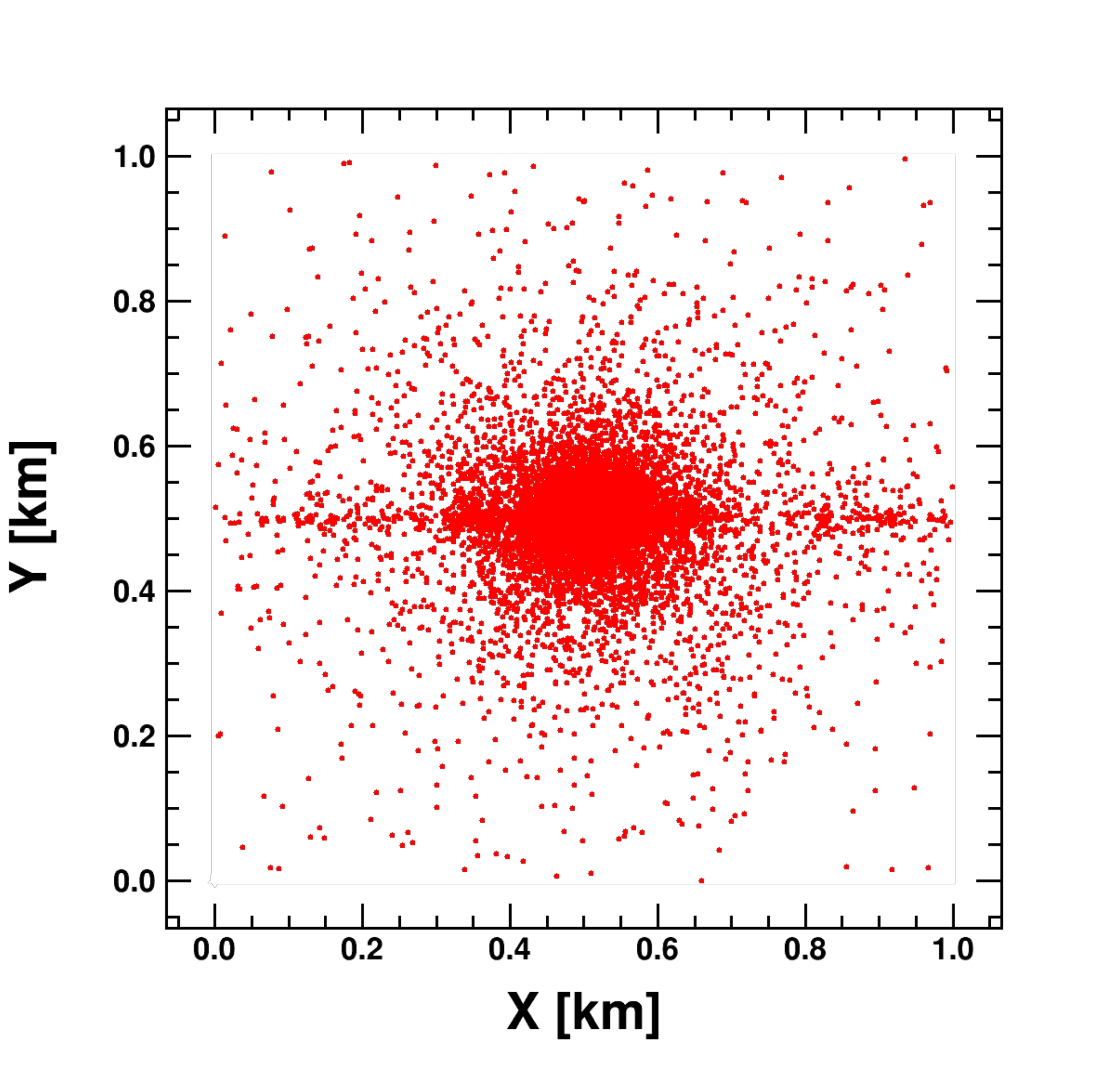}& \includegraphics[width=0.45\textwidth]{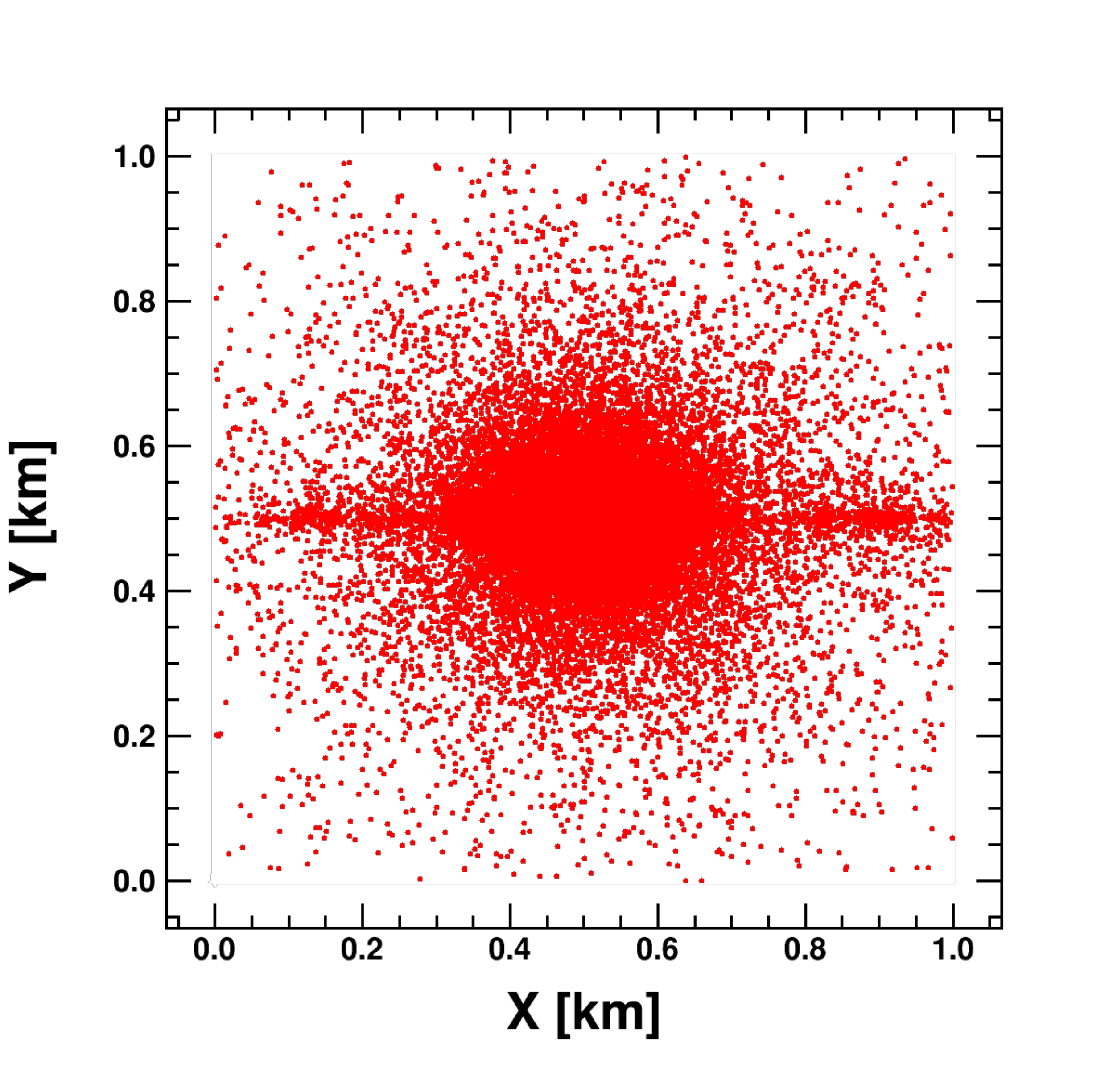}
\end{array}$
\end{center} 
\caption{Detected footprint of an SPS generated by a $10^{19}$ eV photon by detector arrays of different geometries, all located on ground. All the figures share the same covered area which corresponds to 1 km$^{2}$ whereas the spacing between detector units is 25 cm. The single unit detector is presented as a square area. \textit{Upper row:} Detector area of 0.16 cm$^{2}$ (left) and 0.25 cm$^{2}$ (right).
\textit{Middle row:} Detector area of 2 cm$^{2}$ (left) and 25 cm$^{2}$ (right).
\textit{Lower row:} Detector area of 100 cm$^{2}$ (left) and 400 cm$^{2}$ (right).}
\label{fig:ground}
\end{figure*}

In order to demonstrate the capabilities of detection of multi air shower footprints  on Earth, we have performed simulations of different geometrical configurations of ideal ground detectors. In order to derive a distribution of the of SPS-induced air shower particles on the surface of the Earth, simulations with the CORSIKA program~\cite{Heck:1998vt} were performed. The multi air shower particle distributions were obtained taking as an input to CORSIKA the spatially extended SPS distributions generated with the modified PRESHOWER program, with several additional adjustments to keep the compatibility with CORSIKA. The resultant particle distributions form very characteristic, "galaxy-shaped" footprints composed of many extended air showers with significantly dispersed cores, as demonstrated with an example shown in Figure~\ref{fig:footprint}.
By applying a simple geometrical study we demonstrate below that the SPS footprints are not only reconstructable, but that they can also be clearly distinguished from particle distributions typical of single EAS. Figure~\ref{fig:ground} qualitatively shows the detection capability of an example SPS footprint shape after applying several detector array configurations featuring variable single detector dimensions and positioning. We assume ideal conditions for detection: particle falling inside the planar box of the detector is detected with $100\%$ efficiency. Within the figure, we keep the detector array size and spacing between individual detector units fixed while varying their sizes. It is clearly seen that the characteristic "galaxy-like" shapes of SPS footprints are reconstructable, although the required detector array parameters might be economically demanding.  But even if a characteristic shape of the central regions of an SPS-induced particle distribution cannot be reconstructed, one can base the experimental strategies on observing air shower "walls": groups of extensive air showers with parallel axes, all practically contained within one plane. The projection of such a plane onto the Earth surface might span the whole hemispheres of the globe and provide very promising experimental opportunities. While a detailed planning of the relevant observational strategies requires a dedicated follow-up study which is still in progress, the qualitative picture presented in this report might serve as an argument in favor of the SPS detection feasibility, also in terms of the expected event rates.
Interestingly, the recent work by~\cite{AlmeidaCheminant:2020sfw} points out at the possibility of several SPS events a year to be seen by a large observatory due to a possible "boosting" caused by gamma-ray bursts seen as point sources with a corresponding estimated flux one order of
magnitude higher than the upper limits determined by the Pierre Auger Observatory and Telescope Array. In fact, we can consider the flux from the Pierre Auger Observatory upper limits of diffuse flux of
UHE photons of energies above 10 EeV, $\phi^{diff}_{\gamma}(10~\textrm{EeV})\sim 2\times 10^{-3}$ km$^{-2}$yr$^{-1}$sr$^{-1}$~\cite{Savina2021} in order to estimate the fraction of events emitted from vicinity of the Sun where an SPS is likely to originate. By considering a ring of solid angle with external radius of $R_{ext}=2.5$R$_{\odot}$ optimal for magnetic pair production as displayed in Figure~\ref{fig:conversion_p}, and internal radius $R_{int}=1$R$_{\odot}$ with value of $3.5*10^{-4}$ sr, we can estimate for the Pierre Auger Observatory, whose effective area is of about 3000 km$^{2}$ and lifetime of about 30 years, a number of events of about $2*10^{-3}\times 3.5*10^{-4} \times 3000 \times 30\sim0.06$. This example event rate tells us about a conservative but already non-negligible chance for a new (unobserved) physics detection with the available infrastructure, if we assume that a specific characteristics (remarkable elongation of particle distribution) of SPSs makes them recognisable under ideal conditions for detection, and that the Pierre Auger Observatory or a detector array of a similar size can be tuned to be sensitive to SPSs with a reasonable efficiency. Furthermore, the expected SPS event rate might grow if a joint, multi-observatory analysis is being performed continuously, if we consider point sources of UHE photons and the corresponding flux limits \cite{TelescopeArray:2020hey,PierreAuger:2016ppv}, if we consider  
possible transient boosting of the emission from point sources of UHE photons, e.g. an amplification of the flux up to 652 times, as discussed in Ref.~\cite{AlmeidaCheminant:2020sfw}, and finally if we consider a possible sensitivity to the groups of particles propagating far from SPS cores, in the tails extending even over many millions of kilometers (as seen in Figs. \ref{fig:signature}, \ref{fig:foot_print_size}, and \ref{fig:foot_print}), as it would increase the solid angle around the Sun from where an observable SPS could be expected.
%
\begin{figure*}[!bpht]
\begin{center}$
\begin{array}{cc}
\includegraphics[width=0.45\textwidth]{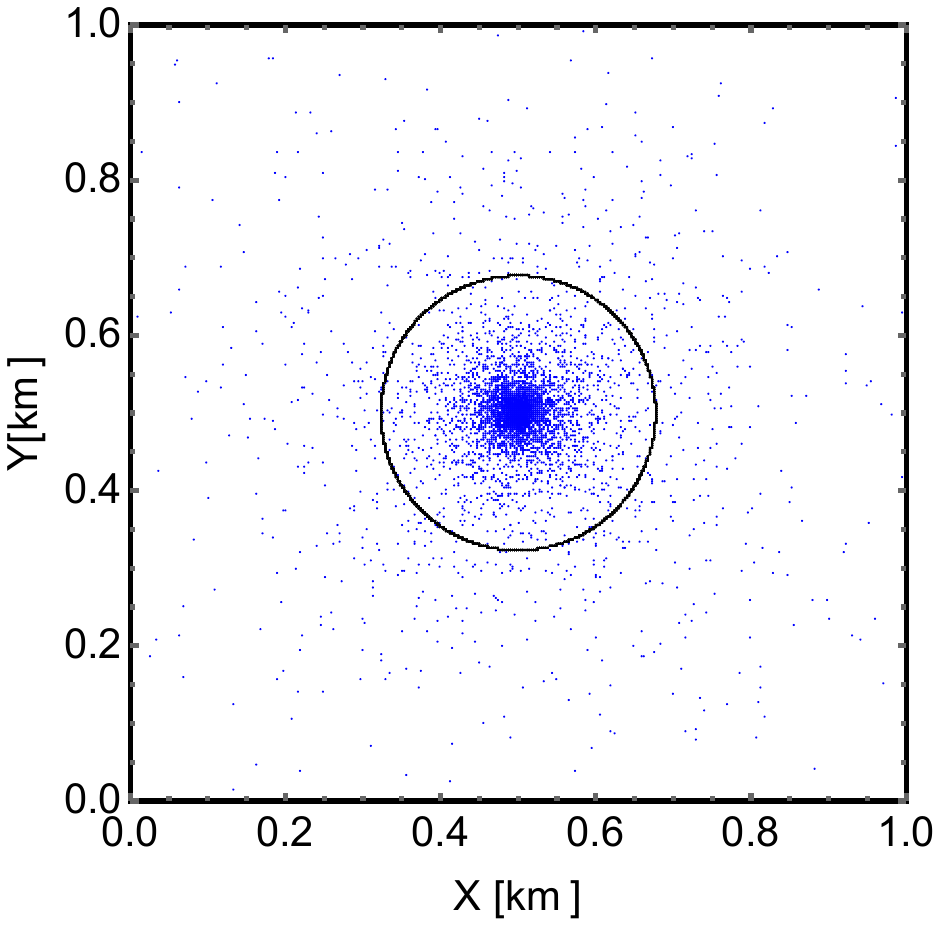} & \includegraphics[width=0.45\textwidth]{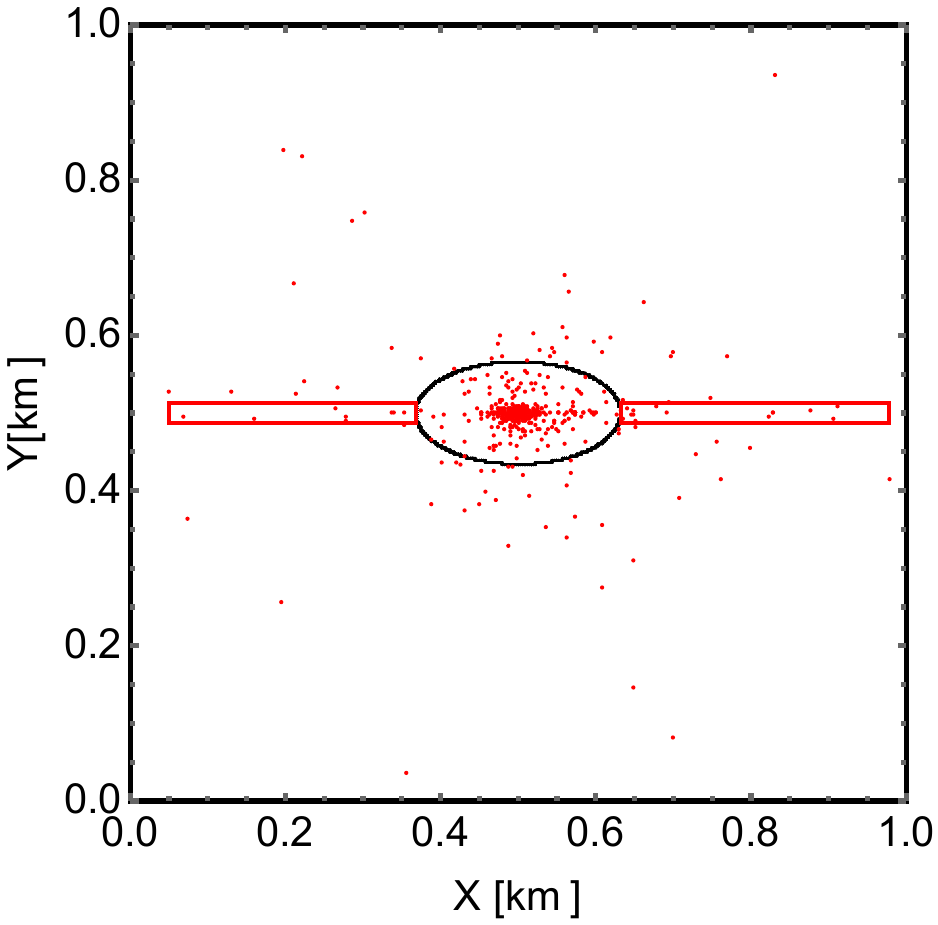}\\
\end{array}$
\end{center} 
\caption{Footprints of an unconverted photon (left panel) and of an SPS (right panel), both of energies of 10 EeV, arriving vertically at the Earth, and detected on ground by an array of sensors, each of them with a collecting area of 2 cm$^{2}$, all spaced  by 25 cm (as in the left panel of the middle row in Fig. \ref{fig:ground}). The areas inside the black contours contain $90\%$ of the particle distributions whereas the red rectangles on the right panel serve to guide the eye in order to highlight the part of the remaining detected particles within a rather extended distribution characteristic of the \textit{galaxy-shape} footprint, as clearly seen in Fig.~\ref{fig:ground}.}
\label{fig:galaxy_plot}
\end{figure*}
A characteristic spatial elongation of SPSs demonstrated above indicates that particle distributions of central parts of SPSs are very peculiar in comparison with single air shower footprints. In order to understand the topology of multi air shower footprints generated by SPSs, we then choose to characterize the corresponding particle distributions by comparing them to those induced by unconverted UHECR photons. The left panel of Figure~\ref{fig:galaxy_plot} presents the case of an individual air shower induced by a primary photon of energy $10^{19}$ eV, arriving vertically at the detector site. Here the black circle denotes the area containing $90\%$ of secondary particles that comprise an air shower. On the other hand, as can be seen in the right panel of Figure 10, the multi air shower footprint produced by an SPS generated by a photon of the same energy has a compact particle distribution around the cores of the central, most energetic showers accompanied by an extended, very thin area containing air showers of lower energies. We conclude that the elongated, "galaxy"-like shapes of the SPS-induced particle distributions on the ground are clearly distinguishable from the footprints of individual extensive air showers, and that the multi air shower particle distribution might potentially be observable under conditions of $100\%$ detection efficiency due to the aforementioned characteristic pattern.

The ongoing follow-up studies dedicated to specific UHE photons scenarios, applying more realistic detector configurations, and involving more precise particle distributions will help to quantify the SPS event rate expectations and detection efficiencies attainable with particular infrastructure capabilities. One of the promising experimental initiatives within which relevant studies and the corresponding experimental efforts are being undertaken is the Cosmic Ray Extremely Distributed Observatory (CREDO) Collaboration \cite{Homola:2020odt}. CREDO aims at the search for large scale cosmic ray correlations using the available and future data on cosmic and gamma ray events of energies that span the whole cosmic ray energy spectrum, and the results presented in this article contribute directly to the CREDO science program.

\section{Summary and Prospects}
Our simulation results show that photons in SPS cascades are extended over a huge spatial extent
(even millions of kilometers) practically along a line. The orientation and size of
these line-like signatures, however, depend on the initial direction and  impact position of the primary UHE photon relative to the solar magnetic field.
Also, photons in SPS cascades can possess energies that
span more or less the whole cosmic-ray energy spectrum, from below GeV to above an EeV.

Detection of SPS cascades is limited by two major requirements. The first is the size of the detector itself, which should be big enough to detect SPS particles
distributed over very large distances. The other obvious requirement is that the detector should be operational during the daytime. As such, only a large array of 
ground-based particle detectors like the Pierre Auger Observatory \cite{2015172} would be suitable for SPS detection. However, the most promising experimental approach to SPS observation and studies should go even further to form a global alliance of all radiation detector arrays and individual sensors capable of detecting secondary particles from the extensive air showers produced by SPSs. Such an alliance is envisaged by the (CREDO) experiment~\cite{Homola:2020odt}.

Given the current UHE photon limits \cite{Savina2021,TelescopeArray:2020hey,PierreAuger:2016ppv}, the expected number of SPSs with cores landing within an observatory of the size of Auger is small, altough non-negiligible. A multi-collaboration SPS observation campaign, the aforementioned possibility of temporarily increasing directional event rates related e.g. to gamma ray bursts, as well as the sensitivity of a detector network to the tails of the significantly elongated SPSs allows expecting an experimentally interesting rate of events arriving from a considerably large region of the sky, not only from a region around the Sun.

SPS-like processes at other sites in the Universe as well as other physics processes might also produce a ``shower'' of correlated particles,
the cosmic ray ensembles (CREs), while they propagate 
in space. Thus, other stellar objects which have a magnetic field strength at least of the order of 0.1 G at their 
surfaces will also initiate SPS-like CREs.
If we assume that a UHE photon undergoes an SPS-like process 
in regions of the Universe with relatively stronger magnetic field while heading towards the Earth and we have a cosmic ray detection framework
that can detect two or more photons simultaneously at very distant locations,  the ``explorable horizon'' for such process can be estimated using simple geometry considerations.
From our SPS simulation results for 100 EeV photon, minimum distances between the most energetic ($ >1$ EeV), and low energy (1 -- 10 TeV)
SPS photons as the cascade reaches the top of the Earth's atmosphere are both of the order of the order of 0.001 m.
Provided a framework which can detect these ``close photons'' in the CREs  arriving as far as \mbox{10 000 km} apart at the Earth from extragalactic regions or sites,
the ``horizon''  is extended to $\sim 100 \mathrm{\ kpc}$, i.e. roughly to the size of our Galaxy.
For comparison, the mean free path for gamma-rays at 1 EeV (1 TeV) is of the order of 100 kpc (500 Mpc).
Photon splitting in strong magnetic fields in the proximity of neutron stars \cite{ADLER1971599, 1997ApJ...476..246H} is another process which is capable of producing CREs,
of which the estimation of the expected signature at the Earth requires a dedicated study and will be performed in the near future.
Although we are not certain about the expected rate of CREs, these, together with SPSs constitute a yet-unchecked scenario that is easy to verify and has a potential of opening
a new window to the Universe.

\begin{acknowledgments}
This work was partly funded by the International Visegrad Fund Grant No. 21720040 and 21920298 and by the National Science Centre Grants No.  2016/23/B/ST9/01635 and 2020/39/B/ST9/01398. This research has also been supported in part by PLGrid Infrastructure. We warmly thank the staff at ACC
Cyfronet AGH-UST, for their always helpful supercomputing support. CREDO mobile application was developed in Cracow University of Technology. D. A-C. acknowledges support from the Bogoliubov-Infeld program for collaboration between JINR and Polish Institutions as well as from the COST actions CA15213 (THOR) and CA16214 (PHAROS). The work of J.Z-S. was funded by ANID-Millennium Science Initiative Program - ICN2019\_044.
\end{acknowledgments}

\bibliographystyle{JHEP}
\providecommand{\href}[2]{#2}\begingroup\raggedright\endgroup
\end{document}